\documentclass[useAMS,usenatbib]{mn2e}
\usepackage{graphicx}
\usepackage{hyperref}
\usepackage{amsmath}

\hypersetup{
  colorlinks=true,
  urlcolor=blue,
  citecolor= blue,
  linkcolor = blue
}

\def\ltsima{$\; \buildrel < \over \sim \;$}
\def\simlt{\lower.5ex\hbox{\ltsima}}
\def\gtsima{$\; \buildrel > \over \sim \;$}
\def\simgt{\lower.5ex\hbox{\gtsima}}

\title[The stellar halo profile with CFIS]{ A-type stars in the Canada-France Imaging Survey I. The stellar halo of the Milky Way traced to large radius by blue horizontal branch stars}
\author[G. F. Thomas et al.]{Guillaume F. Thomas$^{1}$\thanks{E-mail:
guillaume.thomas@nrc-cnrc.gc.ca}, Alan W. McConnachie$^{1}$, Rodrigo A. Ibata$^{2}$, Patrick C\^ot\'e$^{1}$,
\newauthor Nicolas Martin$^{2,3}$ , Else Starkenburg$^{4}$, Raymond Carlberg$^5$, Scott Chapman$^6$,
\newauthor S\'ebastien Fabbro$^{1}$, Benoit Famaey$^{2}$, Nicholas Fantin$^{1}$, Stephen Gwyn$^{1}$,
\newauthor Vincent H\'enault-Brunet$^{1}$, Khyati Malhan$^{2}$, Julio Navarro$^7$, Annie C. Robin$^8$,
\newauthor Douglas Scott$^9$
\\$^{1}$NRC Herzberg Astronomy and Astrophysics, 5071 West Saanich Road, Victoria, BC, V9E 2E7, Canada\\$^{2}$Universit\'e de Strasbourg, CNRS, Observatoire astronomique de Strasbourg, UMR 7550, F-67000 Strasbourg, France\\$^{3}$Max-Planck-Institut f\"ur Astronomie, K\"onigstuhl 17, 69117 Heidelberg, Germany\\$^{4}$Leibniz Institute for Astrophysics Potsdam (AIP), An der Sternwarte 16, D-14482 Potsdam, Germany\\$^{5}$Departement of Astronomy and Astrophysics, University of Toronto, Toronto, ON M5S 3H4, Canada\\$^{6}$Department of Physics and Atmospheric Science, Dalhousie University, Coburg Road, Halifax, NS B3H 1A6, Canada\\$^{7}$ Departement of Physics and Astronomy, University of Victoria, Victoria, BC, V8P 1A1, Canada \\$^{8}$Institut UTINAM, CNRS UMR6213, Univ. Bourgogne Franche-Comt\'e, OSU THETA Franche-Compt\'e-Bourgogne,\\Observatoire de Besan\c{c}on, BP 1615, 25010 Besan\c{c}on Cedex, France\\$^{9}$Dept. of Physics and Astronomy, University of British Columbia, Vancouver, B.C., V6T 1Z1, Canada}

\date{Accepted September 19th 2018}

\begin{document}

\pagerange{\pageref{firstpage}--\pageref{lastpage}} \pubyear{2018}

\maketitle

\label{firstpage}

\begin{abstract}
We present the stellar density profile of the outer halo of the Galaxy traced over a range of Galactocentric radii from $15< R_{GC} < 220$ kpc by blue horizontal branch (BHB) stars. These stars are identified photometrically using deep $u-$band imaging from the new Canada-France-Imaging-Survey (CFIS) that reaches 24.5 mag. This is combined with $griz$ bands from Pan-STARRS 1 and covers a total of $\sim4000$ deg$^2$ of the northern sky. We present a new method to select BHB stars that has low contamination from blue stragglers and high completeness. We use this sample to measure and parameterize the three dimensional density profile of the outer stellar halo. We fit the profile using (i) a simple power-law with a constant flattening (ii) a flattening that varies as a function of Galactocentric radius (iii) a broken power law profile. We find that outer stellar halo traced by the BHB is well modelled by a broken power law with a constant flattening of $q=0.86 \pm 0.02$, with an inner slope of $\gamma=4.24 \pm 0.08$. This is much steeper than the preferred outer profile that has a slope of $\beta=3.21\pm 0.07$ after a break radius of $r_b=41.4^{+2.5}_{-2.4}$ kpc. The outer profile of the stellar halo trace by BHB stars is shallower than that recently measured using RR Lyrae, a surprising result given the broad similarity of the ages of these stellar populations.
\end{abstract}


\section{Introduction}

 \begin{figure*}
\centering
  \includegraphics[angle=0, clip, width=12cm]{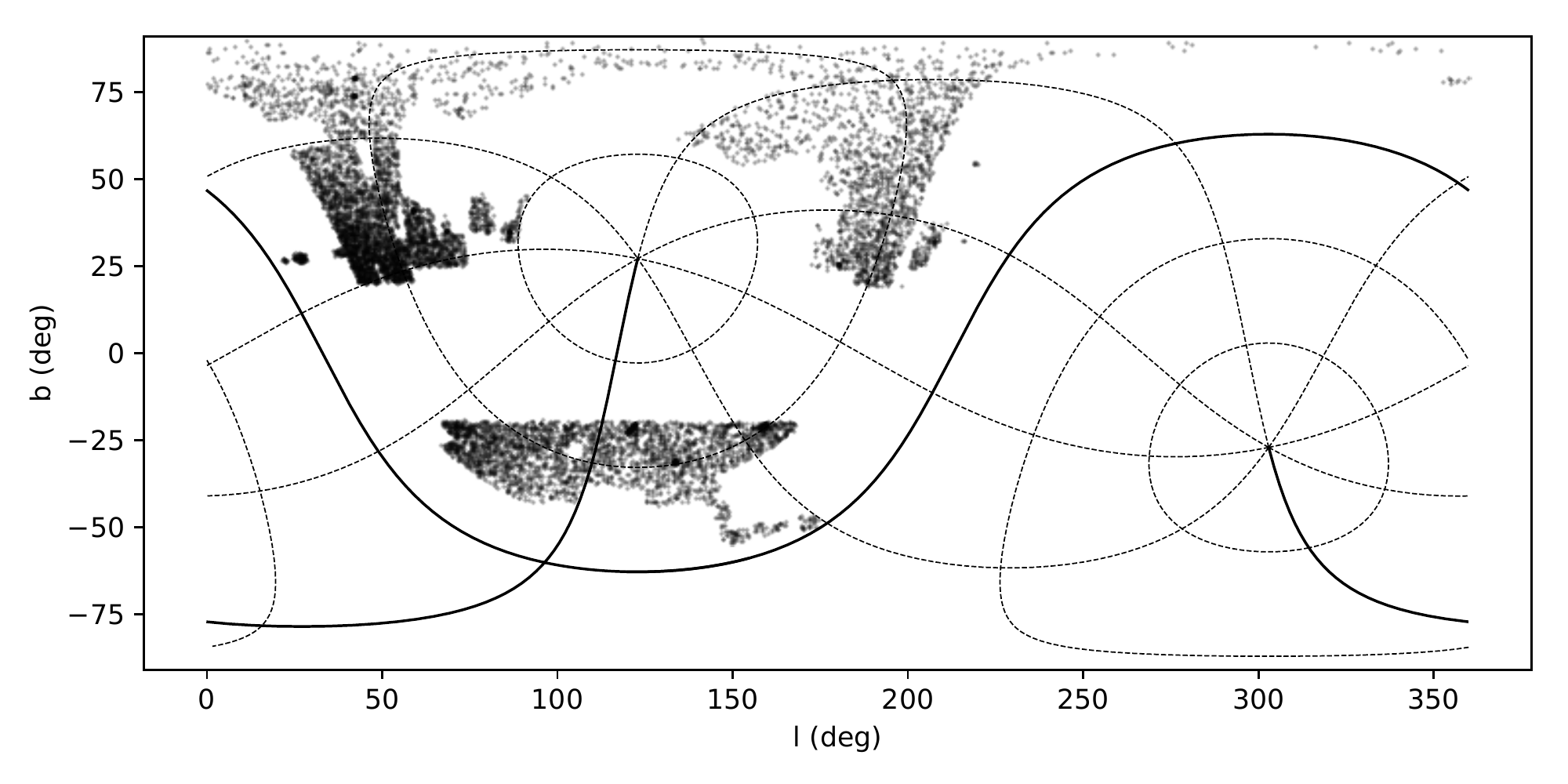}
   \caption{The spatial coverage of CFIS at the time of our study. Specifically, the grey points show the apparent position of the BHBs from our sample. The black lines show the equatorial coordinates, with the equatorial plane highlighted with a solid line.}
\label{plot_lb}
\end{figure*}

It is now generally accepted that large galaxies, like the Milky Way, have been formed by a succession of mergers and via the accretion of smaller galaxies, in a process called hierarchical formation. In the case of accretions, the smaller galaxy is disrupted due to the tidal effects generated by the larger (host) galaxy. This leads to the formation of stellar streams clearly visible around many massive galaxies of the Local Group \citep[e.g.][]{martinez-delgado_2010,martin_2013,grillmair_2016,bernard_2016,malhan_2018}. Although these structures stay spatially coherent for many Gyr \citep{johnston_2008}, they tend to be eventually destroyed by mixing effects and are in turn assimilated to form part of the ``smooth'' stellar halo.

The stellar halo of a $L\star$ galaxy can be a complex structure, very inhomogeneous and clumpy. Nevertheless, it is possible to view it as a smooth component with halo substructures, from which we can study the accretion history, in particular of the Milky Way. Indeed, as shown in many cosmological simulations \citep{bullock_2005,abadi_2006,johnston_2008,cooper_2010,pillepich_2014,pillepich_2018,amorisco_2017}, the accretion history of a galaxy has a huge impact on the profile of the smooth stellar halo component, such that galaxies having the most quiescent accretion history tend to have a profile less steep than for a galaxy of the same mass that has had a much more active accretion history \citep{libeskind_2011}. Most studies of the profile of the Milky Way stellar halo do not go beyond $\sim 100$ kpc, due to the depth of large surveys like the Sloan Digital Sky Survey (SDSS) and the faint absolute magnitude of tracers \citep{yanny_2000,bell_2008,watkins_2009,depropris_2010,pila-diez_2015,xue_2015,slater_2016}. Large area surveys are essential in order to gain constraints on the three dimensional shape of the stellar halo \citep{cohen_2017,fukushima_2018}. 

Blue Horizontal Branch stars (BHB) are ideal tracers for studying the profile of the outer stellar halo ($R_{GC} > 20$ kpc). They are present in old stellar populations, and their absolute magnitude is roughly constant and bright $M_g \simeq 0.5$ \citep{deason_2011}, meaning that they can be identified even at very large distances ($> 100$ kpc).  SDSS, although covering a large portion of the sky ($\sim 14,000$ deg$^2$), does not have a deep enough $u$-band to study the halo beyond 100 kpc \citep{deason_2014}. The new $u$-band coverage provided by the Canada-France-Imaging-Survey (CFIS), intended to eventually cover $\sim 10,000$ deg$^2$ of the northern hemisphere, is $\sim 2.5$ magnitudes deeper than the $u$-band of the SDSS \citep{ibata_2017a}. It is therefore now possible to study the stellar halo for a large fraction of the sky up to a galactocentric distance of $\sim 220$ kpc with BHB stars identified in CFIS.

In this article, we study the three-dimensional profile of the outer stellar halo with a sample of BHB stars, selected through their photometry using the CFIS and Pan-STARRS 1 data that we will present in Section~\ref{data}. Section~\ref{BHBsel_par} presents a new method to disentangle BHB stars from other stellar populations, especially the Blue Stragglers (BS), and we determine the distances to our BHB sample (Section~\ref{distance}). Then, in Section~\ref{sec_comp}, we present our study of the completeness of the BHB sample, including spatial variations. Section~\ref{stellar-profil} describes our derivation of the radial profile and its parameterization. Then, in Section~\ref{results}, we discuss the results and compare our best-fit parameters with those found in previous work. Finally we summarize our results in Section~\ref{conclusions}. 

\section{Data} \label{data}

 \begin{figure*}
\centering
  \includegraphics[angle=0, clip, width=16.0cm]{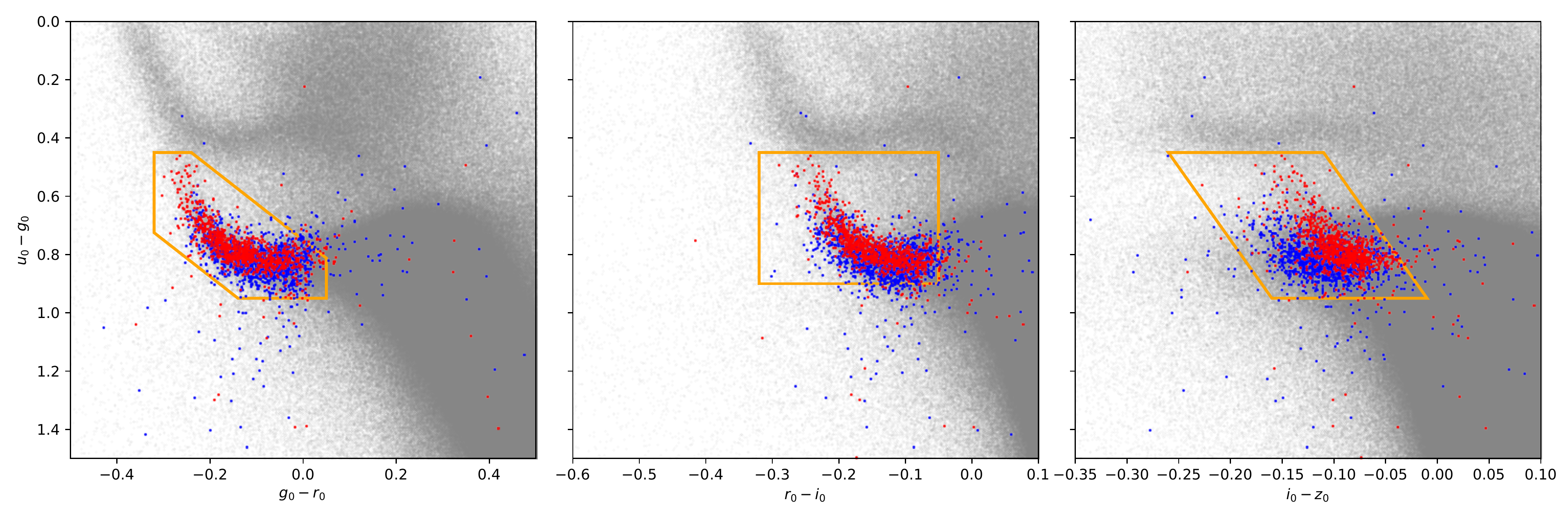}
   \caption{Stellar colour-colour diagram where the grey dots are point sources in the CFIS-PS1 data and the red and blue dots are respectively the BHB and BS samples of \citet{xue_2011} . The orange boxes show the different colour cuts that we use to select A-type stars.}
\label{colourcolour}
\end{figure*}

The primary source of observational data used in this study is a merged catalogue using the \textit{griz} bands from Pan-STARRS 1 \citep{chambers_2016} (hereafter, PS1; specifically, we use the forced PSF photometry parameters) and the \textit{u}-band from CFIS \citep{ibata_2017a}. The PS1 survey covers more than $3/5$ of the sky; as such, the spatial coverage of our merged catalog is limited by the spatial coverage of the CFIS data at the time of our study ($\sim 4,000$ deg$^2$). CFIS excludes most of the Galactic disk by applying a cut in Galactic latitude at $b < 19 $ deg, and the current footprint is limited to a declination of $\delta \le 60$ deg. The current footprint of the CFIS survey is visible in Figure~\ref{plot_lb}.

By cross-matching the CFIS and PS1 catalogs, we retain 98\% of the PS1 detections in the relevant magnitude range. 

Following \citet{farrow_2014}, we use the following criterion to separate stars from the background galaxies, defined in the PS1 $i-$band:
\begin{equation}
i_{PSF}-i_{Kron} < 0.05 \,.
\label{star1}
\end{equation}

It is worth noting that star -- galaxy classification done in this way becomes unreliable at $i_{PSF} \gse 21 $; nevertheless, more than 93\% of the final sample of BHB stars extracted in this survey have $i<21$, thus star -- galaxy misclassification is not expected to have a large impact on the results of this study.

To correct for the Galactic foreground extinction, we used the extinction values, $E(B-V)$, of \citet{schlegel_1998},  assuming the conversion factor given by \citet{schlafly_2011} for a reddening parameter $R_v=3.1$. For the $u$-band  of the CFIS survey, we have assumed that this coefficient is approximately the same as the coefficient of the $u$-band for SDSS. 
We limit our dataset to have photometric uncertainties $<0.2$ mag in each bands ($u$, $g$, $r$, $i$ and $z$).

In what follows, we used the spectroscopic sample of A-stars from \citet{xue_2011}, mostly composed of BHB and Blue Stragglers, cross-matched to CFIS, as the training set for the Principal Component Analysis (PCA) described below.
 
\section{The CFIS BHB stars} \label{BHBsel_par}

\subsection{Selection of BHB stars}

BHBs are hot, A-type, stars (7500 $\lse$ T$_{eff}$ $\lse$ 9000 K). A-type stars can be easily identified and separated from others types of stars with colour-colour cuts involving the $u$-band \citep{yanny_2000, sirko_2004,deason_2011}. However, we note that using only the $(g_0-r_0)$ vs $(u_0-g_0)$ colour-colour diagram to select A-stars, like \citep{sirko_2004,deason_2011},  results in significant contamination from cooler stars. 

We select A-type stars using three different colour-colour diagrams ($(g_0-r_0)$ vs $(u_0-g_0)$,  $(r_0-i_0)$ vs $(u_0-g_0)$ and $(i_0-z_0)$ vs $(u_0-g_0)$) as shown in Figure \ref{colourcolour}, where the red dots are the spectroscopic  BHB sample of \citet{xue_2011}. It is important to note that the $(u_0-g_0$) colour of our catalog is shifted by $\simeq$ 0.3 mag compared to the same colour using the SDSS $u$ and $g$ filters, since the filters are not the same. Applying these selections on the $9.2 \times 10^7$ sources of the cross-matched CFIS-PS1 catalog leads to a sample of $\simeq 29,700$ A-type stars.

Our simple colour cuts select both BHB and BS stars. The latter population have a higher surface gravity than BHB stars ($\log(g)_{BS} \simeq 4.2$ and $\log(g)_{BS} \simeq 3.2$; \citealt{vickers_2012}). This difference in surface gravity between these two populations leads to a difference in the width and the depth of surface gravity sensitive absorption lines such as the Balmer lines around 365 nm and, to a lesser extent, the Paschen lines around 870 nm. This behaviour can be used to disentangle the two populations \citep{sirko_2004,xue_2008}. Indeed, even in the absence of spectroscopic data it is possible to use photometry to discriminate between the BHB and BS. \cite{yanny_2000,sirko_2004,bell_2010} have used the $u$-band and its sensitivity to the Balmer jump to this end, and \citet{lenz_1998} have found that the $i-z$ color is also sensitive to the surface gravity for A-types stars, due to the presence of the Pashen absorption lines in the $z$-band \citep{vickers_2012}. Attempts to use the $u$-band without the $z$-band, or {\it vice-versa}, to disentangle these two populations has generally produced samples of BHB stars that are only $\sim 55 \%$ complete while containing up to $30 \%$ contamination \citep{bell_2010,vickers_2012}. 

As we can see in Figure \ref{colourcolour}, even with these differences between the two populations, it is very difficult to discriminate between them with simple colour-colour cuts. Instead, we try to use all the information available in all the bands. To this end, we developed a discrete classification algorithm using a Principal Component Analysis (PCA) approach based on the work of \citet{ibata_1997a}, where the inputs are the colours $(u_0-g_0)$, $(g_0-r_0)$, $(r_0-i_0)$ and $(i_0-z_0)$. We use the spectroscopic catalog of A-type stars selected by \citet{xue_2011} as the training set to find the principal components that provide the best separation between BHB and BS stars. After cross matching, our training set is composed of $872$ BHB (39.0\% of the training sample) and $1366$ BS (61\%). Following  \citet{ibata_1997a}, we subtracted the mean value from each colour, since this does not contain any fundamental information and avoids the problem of the domination of the covariance matrix by the mean colour (the mean of each colour in our training set is listed in Table \ref{mean_colour}).

\begin{table}
 \centering
  \caption{Mean colours of the training set of A-type stars from \citet{xue_2011}.}
  \label{mean_colour}
  \begin{tabular}{@{}cc@{}}
  \hline
   colour & $\langle$colour$\rangle$ \\
    \hline
  $(u_0-g_0)$ & 0.7970 \\
  $(g_0-r_0)$ & -0.1138 \\
  $(r_0-i_0)$ & -0.1413 \\
  $(i_0-z_0)$ & -0.1050 \\
\hline
\end{tabular}
\end{table}

 \begin{figure}
\centering
  \includegraphics[angle=0, clip, width=7.5cm]{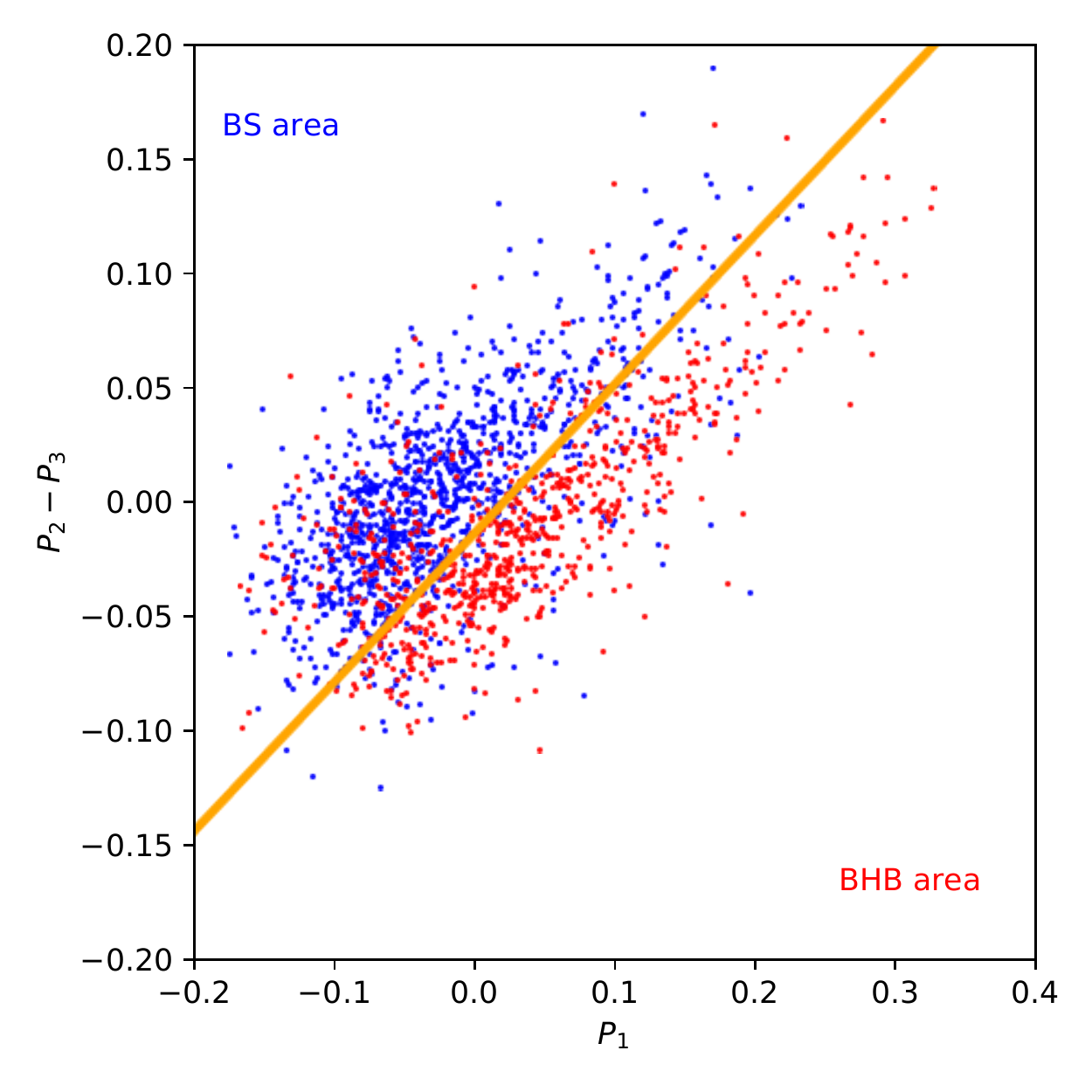}
   \caption{Separation of the BHB and BS area with the two axes determined by the PCA analysis. The red and blue dots correspond, respectively, to BHB and BS from the catalog of \citet{xue_2011}.}
\label{BHBarea}
\end{figure}

The principal components that provide the best separation of the two populations can be found by this algorithm and is described by the following equation, where $P_1$ is the principal component corresponding to the highest eigenvalue of the covariance matrix: 

\begin{equation}
\begin{pmatrix}
P_1 \\ P_2 \\ P_3 \\ P_4
\end{pmatrix}
=
A
\cdot 
\begin{pmatrix}
u_0 - g_0 \\ g_0 -r_0 \\ r_0 - i_0 \\ i_0 - z_0
\end{pmatrix} \, ,
\end{equation}
where 
\begin{equation}
A= \begin{pmatrix}

 -0.6397 & -0.7669  & \quad 0.0493 &  -0.0149  \\
 -0.6479 &  \quad 0.5353 &  -0.2283 & -0.4916 \\
  -0.3964  &   \quad 0.3141 & \quad 0.0040 & \quad 0.8626\\
 -0.1181 &  \quad 0.1633 & \quad 0.9723 &  -0.1183\\

 \end{pmatrix} \
 \end{equation}

We find that the minor axis ($P_4$) does not help to disentangle the two populations. This is unsurprising, since we can see that the $P_4$ axis is mostly influenced by ($r_0 - i_0$), a colour that does not contain hydrogen lines sensitive to the surface gravity that can help to separate the two populations (and which does not play an important role in the construction of the three other axes). Our findings are in line with the idea of \citet{lenz_1998} that the BHB and the BS are separated efficiently using the Balmer and Paschen lines present in the $u$, $g$, $i$ and $z$ bands. In what follows, we just use $P_1$, $P_2$ and $P_3$ to separate the BHB from the BS stars. 

Using these three axes, it is possible to define a region mostly populated by the BHB stars, as seen in Figure \ref{BHBarea}, such that:
\begin{equation}
(P_2-P_3)_{BHB} \le  -0.0141 + 0.6512\, P_1
\label{selBHB}
\end{equation}

Using this definition, we can define a photometric sample of BHB that contains 71\% of the overall BHB sample of the training set, with a contamination from the BS in the training set that is only 24\% of the photometric BHB sample. In order to account for the photometric uncertainties in each band, we have resampled the input colours of the training set 100 times accounting for their measurement errors and we found that, even with the photometric uncertainties, the completeness of the photometric BHB did not change and that the contamination never increased to larger than 26 \% of this sample. 

\begin{figure*}
\centering
  \includegraphics[angle=0, clip, width=16.0cm]{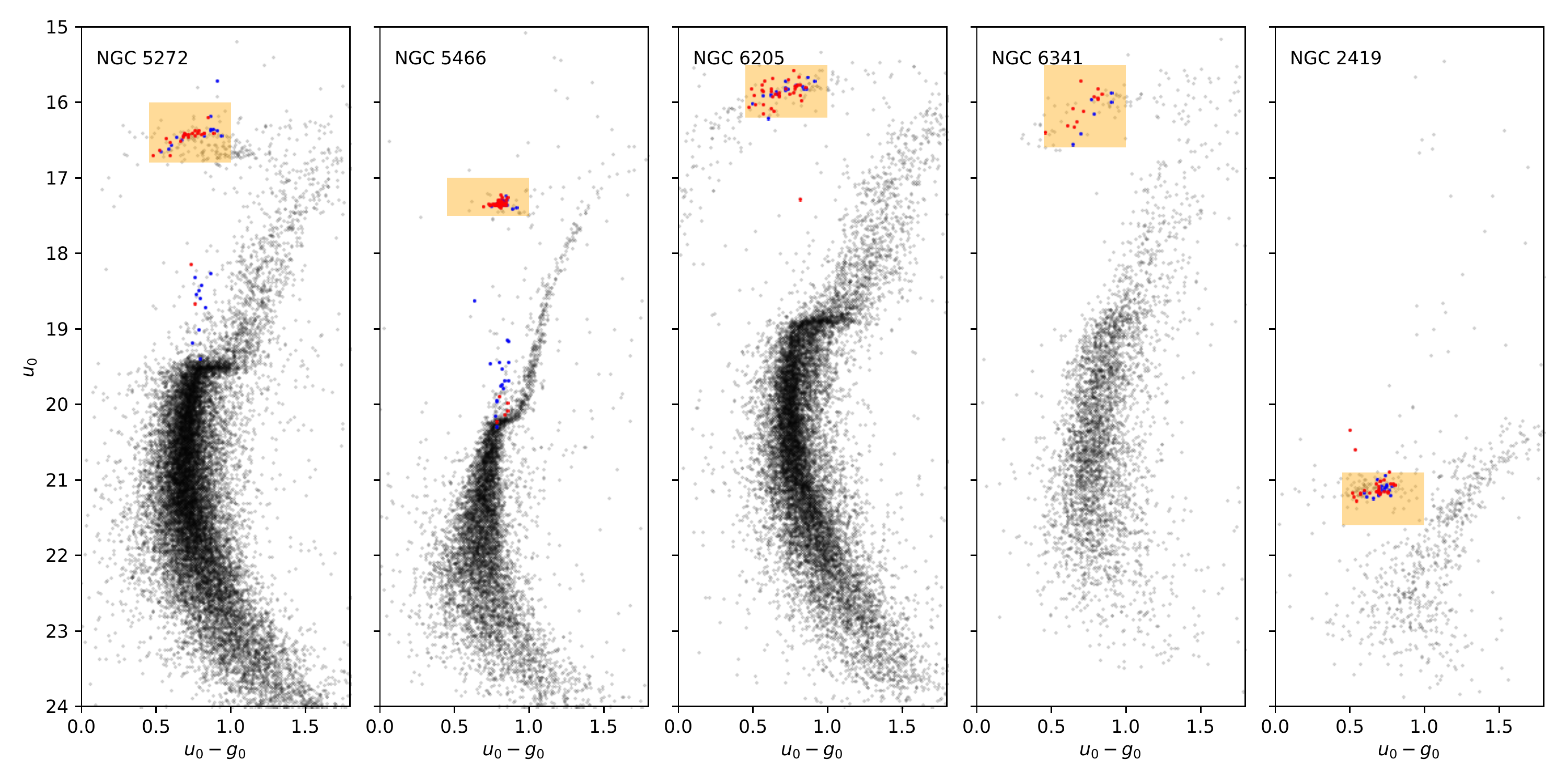}
   \caption{CMDs of five of the seven globular clusters present in the CFIS footprint where the red and blue dots correspond respectively to the stars identify as BHB and BS we our method. The orange area represent the region of the CMD where really lies the BHBs.}
\label{CMD_gc}
\end{figure*}

Seven globular clusters fall within in the current CFIS footprint, NGC 2419, NGC 5272, NGC 5466, NGC 6205, NGC 6341, Palomar 4 and KO 2. However, the latter does not contain any BHB stars due to its very low luminosity of $M_V \sim -1$ mag \citep{koposov_2007}, and Palomar 4 contain only 2 stars that we identify as A-types stars. These 2 clusters are not used in what follows. Using the colour magnitude diagram (CMD) of the other five globular clusters, presented in Figure \ref{CMD_gc}, we visually selected boxes enclosing BHBs in these objects in the $0.45 \le (u-g)_0 \le 0.95$ range. These selection boxes, in orange on Figure \ref{CMD_gc}, contain the stars that we consider to be bonafide BHBs, and which can be used to provide an independent test of the effectiveness of our algorithm.

The resulting completeness and purity of our photometric BHB sample as measured using these globular clusters are shown in Table \ref{purity_GC}. The completeness estimate for each globular cluster is comparable to our earlier estimate using the spectroscopic sample of \citet{xue_2011}. Our method successfully discriminates between the  BHB and the BS in these globular clusters. Indeed, the purity of the BHB is $>90\,\%$  for all these clusters, much higher than for the spectroscopic sample. This high degree of purity is found even for NGC 5272 and NGC 5466, which have the largest populations of BS as inspection of Figure \ref{CMD_gc} makes clear. However, the globular clusters just represent a tiny fraction of the overall population of the stellar halo and may undersample the BS population. Therefore, we adopt the more conservative contamination estimate of 24\% for this study. This contamination rate is similar to those found by \citet{bell_2010} and \citet{vickers_2012}; however, the completeness of our sample is $1.25$ higher than their corresponding BHB samples. 

\begin{table}
 \centering
  \caption{Purity and completeness of our photometric BHB sample in five over seven globular clusters present in the CFIS footprint.}
  \label{purity_GC}
  \begin{tabular}{@{}lccc@{}}
  \hline
   Name & Completeness & Purity  \\
    \hline
  NGC 2419 & 0.63 & 0.93 \\
  NGC 5272 & 0.74 & 0.90 \\
  NGC 5466 & 0.64 & 0.92 \\
  NGC 6205 & 0.68  & 0.94 \\
  NGC 6341 & 0.70  & 1.0  \\
\hline
{\bf Total} & {\bf 0.68} & {\bf 0.94}
\end{tabular}
\end{table}

We calculate the $P_1$ and $P_2-P_3$ axis for all of the $\simeq 29,700$ A-types stars present in the CFIS footprint and selected a photometric sample of BHB stars using Eq.~(\ref{selBHB}). This leads to a photometric BHB sample of $\sim 10,200$ stars. The position of the BHB of our sample in Galactic coordinates is shown in Figure.~\ref{plot_lb}.

\subsection{Distances estimates} \label{distance}

To determine the heliocentric distance of the BHB stars, we use the calibration of the absolute magnitude in the $g$-band ($M_g$) provided by \citet{deason_2011}, which is a function of $(g_0-r_0)$:
\begin{equation}
\small
\begin{split}
M_g =\, &\, 0.434 - 0.169\, (g_{0,} - r_{0})_{SDSS} + 2.319 \,(g_{0,} - r_{0})_{SDSS}^2 \\
&+ 20.449 \,(g_{0,} - r_{0})_{SDSS}^3 + 94.517 \, (g_{0,} - r_{0})_{SDSS}^4 
\end{split}
\normalsize
\end{equation} 

\begin{figure}
\centering
  \includegraphics[angle=0, clip, width=7.5cm]{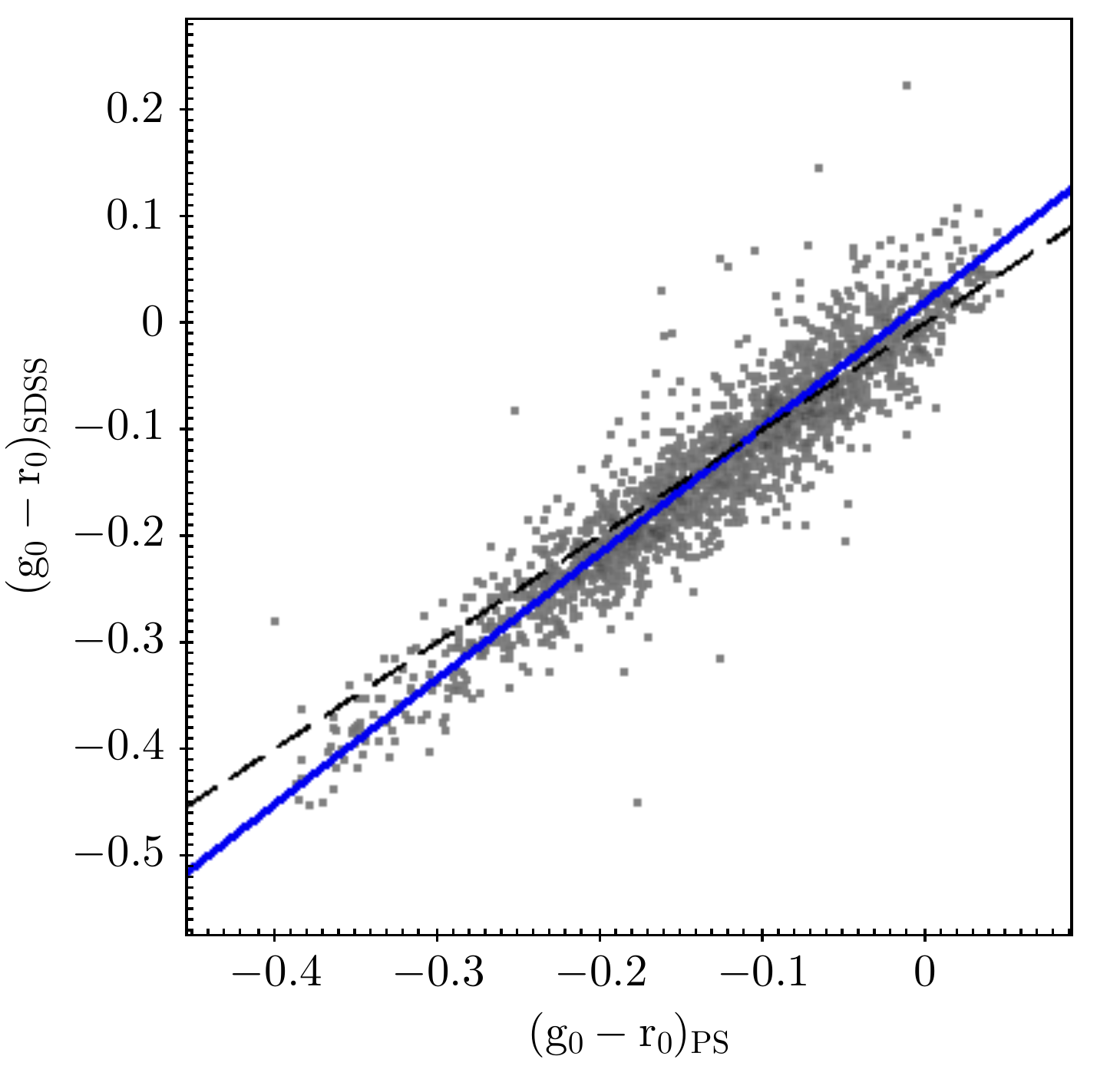}
   \caption{Relationship between the $(g_0-r_0)$ colour using the PS1 and the SDSS filters system. The dashed line shows the relation if $(g_{0,} - r_{0})_{SDSS} = (g_{0,} - r_{0})_{PS}$, and the blue line shows the best fit linear relation described in Eq.~\ref{fit_gr}. }
\label{gr_fit}
\end{figure}

As illustrated in Figure~\ref{gr_fit}, the $(g_0-r_0)$ colour in the Pan-STARRS 1 photometric system is slightly different from the one in the SDSS system, thus we have transformed the $(g_0-r_0)$ colour used in the equation into the Pan-STARRS 1 photometric system by identifying a sub-sample of our A-types stars also identified in the SDSS data release 14. The transformation that we calculate with 1042 stars in this way is given by:
\begin{equation}
(g_{0,} - r_{0})_{SDSS} = 1.18 \,\, (g_{0,} - r_{0})_{PS} +0.02 \, .
\label{fit_gr}
\end{equation} 
The typical difference between the real SDSS colour and that provided by this transformation is 0.01 mag.

Finally, we verify the accuracy of the BHB distances derived in this way by calculating the average distance of the BHBs in our sample that are spatially coincident with five known globular clusters and the Draco dwarf galaxy. These are listed in Table~\ref{dist_GC} and are consistent with the literature values for the distances to these objects. 

\begin{table}
 \centering
  \caption{Comparison of the heliocentric distance of six stellar halo objects derived using the mean magnitude of the BHB ($r_{h, BHB}$ ) with the previous distances to these objects derived using other tracers $r_{h, past}$.}
  \label{dist_GC}
  \begin{tabular}{@{}lccc@{}}
  \hline
   Name & $r_{h, BHB}$ (kpc) & $r_{h, past}$ (kpc) & Source \\
    \hline
  NGC 2419 & 90.8 $\pm 7.8$ & $82.6^{+2.4}_{-1.4}$& \citet{harris_1996} \\
  NGC 5272 & 10.32 $\pm 0.76$ & $10.2 \pm 0.2$ & \citet{harris_1996} \\
  NGC 5466 & 16.13 $\pm 0.25$ & $16.0 \pm 0.4$ & \citet{harris_1996} \\
  NGC 6205 &   7.64 $\pm 0.60$ &   $7.1 \pm 0.2$ & \citet{harris_1996} \\
  NGC 6341 &   8.58 $\pm 0.68$ &   $8.3 \pm 0.2$ & \citet{harris_1996} \\
  Draco dSph &   82.0 $\pm 4.5$ &  79.79 $\pm 2.31$ & \citet{sesar_2017} \\
\hline
\end{tabular}
\end{table}

\section{Evaluation of the completeness} \label{sec_comp}

The most distant BHB star in our sample has a heliocentric distance of $\simeq 220$ kpc. Of course, the fraction of BHBs detected at different distances in our sample depends of the completeness of the survey. The completeness is a function of magnitude, and this in turn varies with position on the sky since, for such a large survey, the depth varies spatially and reflects the specific observational conditions at each position. In this section, we first describe the method that we used to determine the completeness of our survey in the different bands for a reference field of $1 \times 1$ deg$^2$. We then present an analysis of the spatial variation of the limiting magnitude per band.

\subsection{The completeness of the reference field}\label{comp_ref}

The band that has the most influence on the completeness of our sample is not the $u$-band since it is considerably deeper than the PS1 data  ($SNR=5$ at $u \sim 24.5$ \citep{ibata_2017a}). All the other bands are shallower by at least $\sim 1$ mag. The maximum difference of the magnitude between the $u$ and the others bands for A-types stars is less than 1 mag. Therefore,  the completeness of our BHB sample is set by the completeness of PS1.

We estimate the completeness of our sample by comparing the number of sources detected as a function of magnitude to a considerably deeper field in similar bandpasses. To this end, we define a reference field of $1 \times 1$ degree taken from the area covered by the  recent data release 1 of the Hyper Suprime-Cam Subaru Strategic Program \citep[hereafter HSC-SSP,][]{aihara_2018}, which is significantly deeper than PS1. Unfortunately, at the time of our study, there is no region that is covered by both the HSC-SSP and CFIS. However, as mentioned before, the completeness of our BHB sample depends primarily on the completeness of PS1, which has full coverage of all the sky visible from Hawaii. Our reference field is centered at R.A. = 245.5 deg and Dec= 43.5 deg, close to the CFIS footprint and in the HSC footprint.

 \begin{figure}
\centering
  \includegraphics[angle=0,  clip, width=7.5cm]{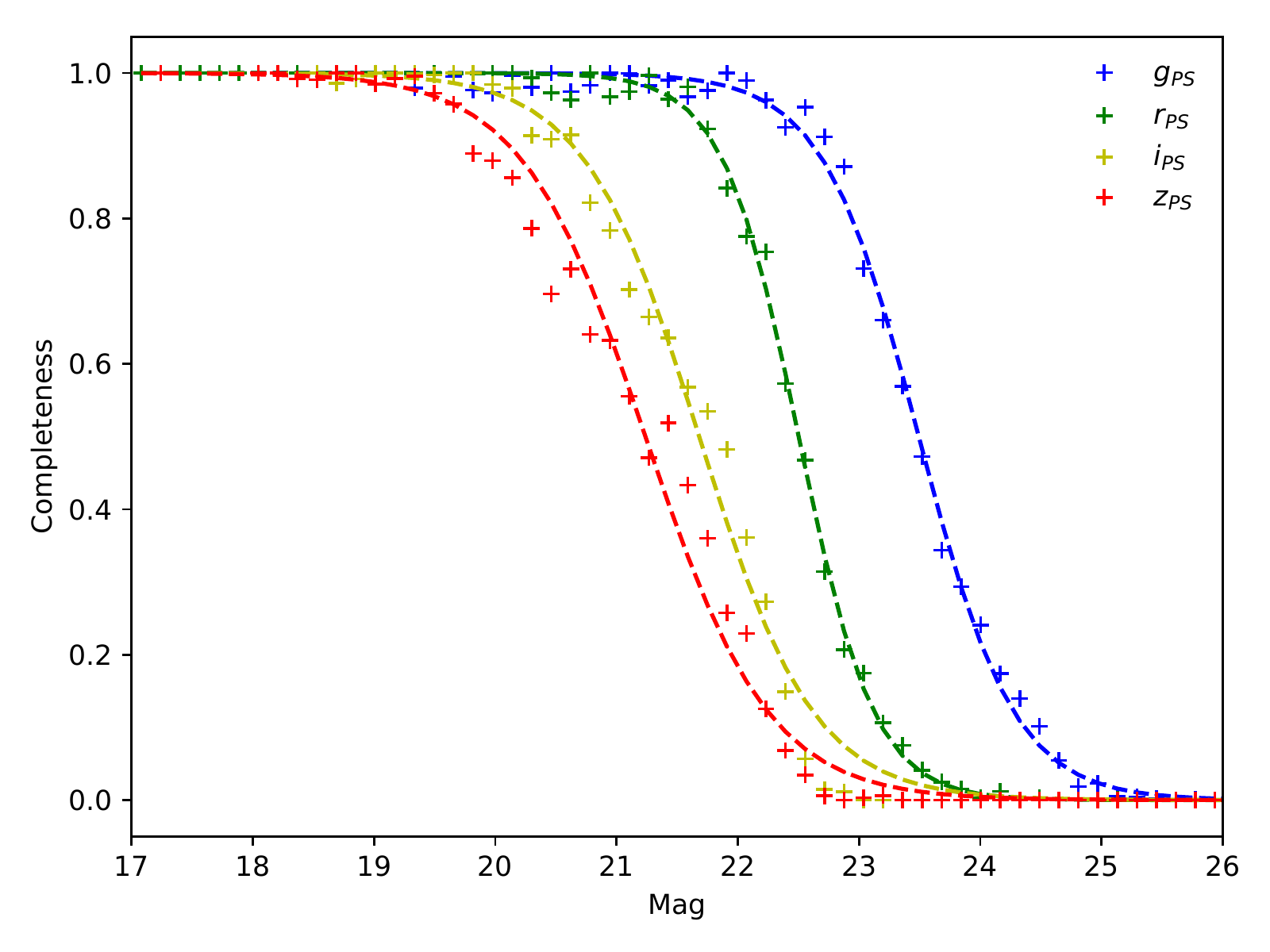}
   \caption{Completeness of the $g$, $r$, $i$ and $z$ bands of the PS1 survey in an area of $1 \time 1$ deg$^2$ centered at (R.A., Dec) = (245.5, 43.5), assuming that all stars brighter than 26 mag are present in the HSC-SSP survey.}
\label{comp}
\end{figure}

\begin{figure*}
\centering
  \includegraphics[angle=0, viewport= 0 95 710 240,  clip, width=18cm]{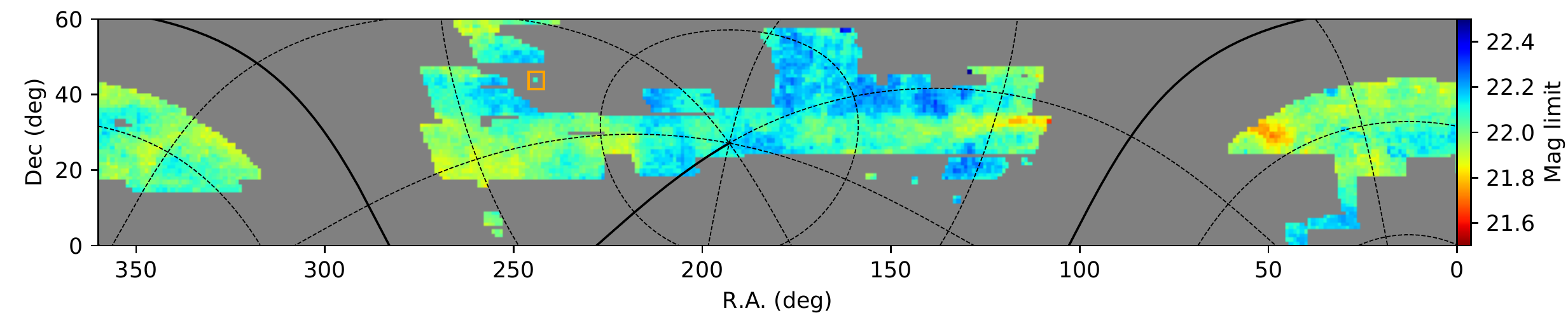}
   \caption{Map of the spatial variation of the limiting magnitude in the $z$-band over the CFIS footprint. The orange square highlights the reference field used to determine the completeness. The black lines show Galactic coordinates, with the Galactic plane and Galactic minor axis highlighted
with a solid line.}
\label{mag_lim}
\end{figure*}

We selected only objects with uncertainties $<0.2$ dex in the $g$, $r$, $i$ and $z$ bands of the two catalogs and applied the criterion defined Equation~(\ref{star1}) on the PS1 data to select only the objects that we identify as stars. We cross-match the catalogs and calculate the fraction of HSC stars\footnote{We defined the stars in the HSC dataset such as $iclassification_{extendedness}=0$.} that are also identified in PS1. The resulting completeness curves for each band are shown in Figure~\ref{comp}, where we have used the following equations to transform the HSC filter system to the PS1 filter system where $k_{RMS}$ is the mean difference between the PS magnitude determine by this equation and the real PS magnitude in the $k$-band :
\begin{equation}
 \left.
  \begin{array}{ l }
g_{PS}=1.005\, g_{HSC} - 0.00025\, g_{HSC}^2 \, \, \, \, g_{RMS}=0.2\\
\\
r_{PS}= r_{HSC} +0.034 \qquad \qquad  \qquad \, \, \, \, \, r_{RMS}=0.11\\
\\
i_{PS}= i_{HSC} +0.1 \qquad \qquad  \qquad   \, \, \, \, \, \, \, \, \, \, \, \,  i_{RMS}=0.08\\
\\
z_{PS}= z_{HSC} +0.1 \qquad \qquad  \qquad \, \, \, \, \, \, \, \, \  z_{RMS}=0.04 \,.
  \end{array}
  \right.
\end{equation}
Although $g_{RMS}$ is large, this imprecision will not have an impact on our study, as the $g$-band is not the band that limits the completeness of our BHB sample (see below).

We find that the data in Figure~\ref{comp} can be reasonably fit with the following generic exponential equation, where $C$ is the completeness in one band:
  \begin{equation}
 C_x = 1.0 /(1.0+ \exp( (x-a)/b) )\, ,
  \label{comp_eq}
\end{equation}
The parameters $a$ and $b$ for each band are listed Table~\ref{param_comp}.

\begin{table}
 \centering
  \caption{ Parameters of the fit of the completeness of $g r i z$ bands of PS1 used in Equation~(\ref{comp_eq}). }
  \label{param_comp}
  \begin{tabular}{@{}ccc@{}}
  \hline
   Band & $a$ & $b$ \\
    \hline
  $g$ & 23.54 & 0.4  \\
  $r$ & 22.55 & 0.31  \\
  $i$ &   21.74 &   0.48 \\
  $z$ &   21.24& 0.51 \\
\hline
\end{tabular}
\end{table}

In Figure~\ref{comp}, it is clear that the $z$-band is the shallowest band, with a $50\%$ of completeness that is $0.45$ mag shallower than for the $i$-band. Moreover, more than $98 \%$ of our sample of A-types stars have a non-dereddened colour $|(i-z)| < 0.2$. Thus we conclude that {\it the completeness of our BHB sample is primarily determined by the completeness of the $z$-band}. Therefore, we use the equation of completeness in the $z$-band in our subsequent analysis to account for completeness effects due to the magnitude limits of the survey (see Section \ref{stellar-profil}). We also note that our reliance on the PS1 data means that we are not yet fully exploiting the depth of the CFIS data, and that we can expect to conduct even deeper studies in the future once deeper $z$-band data become available.

\subsection{Spatial variation of the completeness}

 The HSC-SSP survey covers only a tiny fraction of PS1, and it is impossible to do a similar analysis on the full PS1 footprint to study the spatial variation of the completeness directly. Nevertheless, it is possible to study the variation in the relative depth of the survey, and so relate this back to the completeness of the reference field, through the luminosity function of each band at a given position.

We allow for the fact that the variation in the depth of PS1 may be extremely complex because of the survey strategy, range of observing conditions, and multiple observations of the same region. Therefore, we cut the PS1 survey into ``pixels'' of $1 \times 1$ degrees in right ascension and declination, and calculate the luminosity function per pixel in each of the $g$, $r$, $i$ and $z$-bands. Due to spherical geometry, the number of stars per pixel in the highest declination regions is significantly lower than close to the equatorial plane. However, our survey is currently limited to  $\delta \leq 60$ degrees, and this issue has a negligible impact on the following analysis (the variation of the number of stars per pixel at high declinations is still lower than the Poissonian uncertainty of the most populated pixel). 

We define the ``limiting magnitude'' of each pixel in a given band as the magnitude where the luminosity function, normalized to the maximum value in each pixel, is equal to 0.5. The variation of the limiting magnitude of each pixel of the $z$-band of the PS1 survey over the CFIS footprint is shown on Figure \ref{mag_lim}. The limiting magnitude of the reference field used above to determine the completeness of PS1 is $z_{lim,ref} =22.09$. The mean limiting magnitude over the full CFIS footprint in the $z$-band is of $\langle z_{lim} \rangle =22.06$ with a standard deviation of $\sigma_{z\,lim} =0.01$ mag. 

We then approximate the completeness of a pixel centered on (R.A., Dec) = ($\alpha$, $\delta$) in the $z$-band using Equation~(\ref{comp_eq}) where $z$ is replaced by $z'(\alpha, \delta)$, defined so that: 
\begin{equation}
z'(\alpha, \delta) =z  - (z_{lim}(\alpha, \delta)-z_{lim,ref}) \, .
\label{lim_mageq}
\end{equation}

\begin{figure}
\centering
  \includegraphics[angle=0,  clip, width=7.5cm]{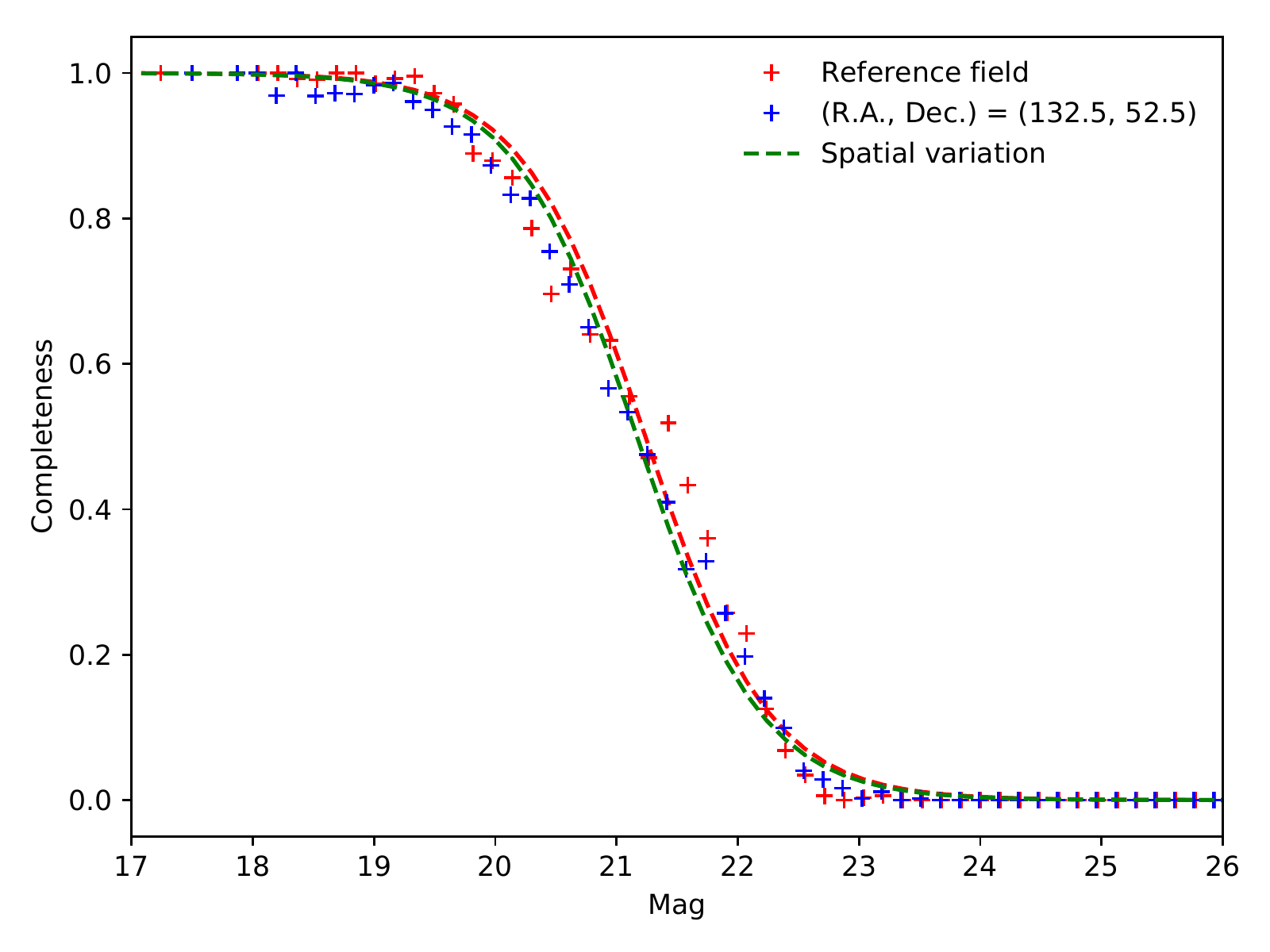}
   \caption{The completeness of the $z$ band of our primary reference field (R.A., Dec) = (245.5, 43.5) is shown in red. The completeness of another field, centered on (R.A., Dec)=($132.5$, $52.5$), is shown in blue, where we have calculated the completeness by direct comparison to HSC data, in the same way as the primary field. The red dashed line is a for for these points. The green dashed line shows the predicted completeness of this field using Eq.(\ref{lim_mageq}) and the method outlined in the text. The two methods agree very well.}
\label{compa_compl}
\end{figure}

We verify that this approximation is valid by comparing the completeness determined in this way with the completeness measured directly (using the technique in the previous section) on a different field covered by the HSC-SSP and PS1. Figure~\ref{compa_compl} shows this comparison for the z-band on a field centered at (R.A., Dec)=($132.5$, $52.5$). It is clear on this figure that this method reproduces very well the completeness of that field: the difference in the z value used to define the two curves in only 0.02 magnitudes. 

 \begin{figure*}
\centering
  \includegraphics[angle=0, clip, width=16.0cm]{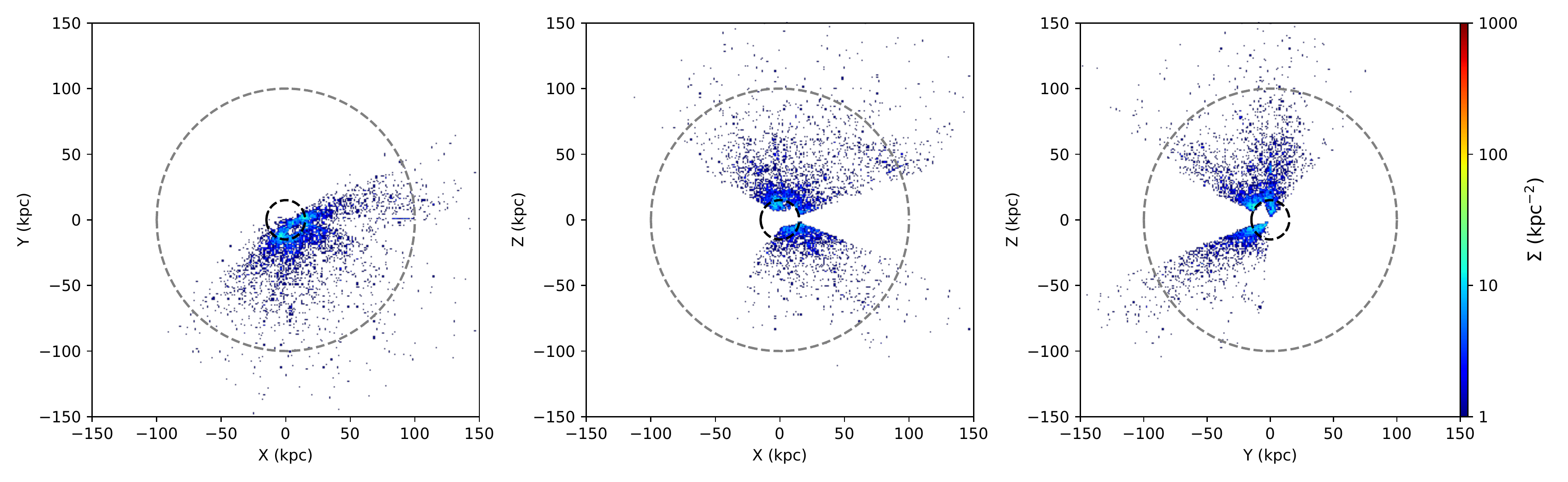}
   \caption{Galactocentric coordinates of the BHB stars. The dark circle correspond to a radius of 15 kpc and the grey circle to a radius of 100 kpc.}
\label{XYZ}
\end{figure*}

\section{The smooth stellar halo profile} \label{stellar-profil} 

In this section, we lay out how we construct a model of the smooth stellar halo traced by our photometric BHB sample, accounting for the observational biases such as the completeness of the sample defined in the previous section. The following analysis is similar to the study done recently with the RR Lyrae of PS1 by \citet{hernitschek_2018}.  However, our selection function, that introduces the observational biases in our model, is slightly different, since the spatial footprint of the surveys and the stellar populations used are different. This kind of approach has been employed by \citet{bovy_2012a} and \citet{rix_2013} for the disk of the Milky Way and by \citet{ibata_2014b} for the stellar halo of the Andromeda galaxy.

In the following, ($X$, $Y$, $Z$) are the Galactic Cartesian coordinates\footnote{In this work, we used the right-hand coordinates, with the $X$ axis pointing toward the Sun and the $Z$ axis toward the North galactic pole.}, $R_{GC} = \sqrt{X^2+Y^2+Z^2}$ is the Galactocentric radius, $r_{helio}$ is the heliocentric distance and $m=\sqrt{X^2+Y^2+(Z/q)^2}$ is the elliptical distance, that allow a vertical deformation of the stellar halo compared to the plane of the Galactic disk through the parameter $q$, such that the stellar halo is spherical if $q = 1$, oblate if $q < 1$ and prolate if $q > 1$. In this work, we assumed that the Sun is located in the plane of the disk ($Z_\odot = 0.0$ kpc) at a distance from the Galactic center of $R_\odot = 8.5$ kpc. {The Cartesian Galactocentric coordinates of the BHB stars is shown in Figure \ref{XYZ}.

\subsection{Stellar distribution model} \label{profile_model}

It is common to model the spatial distribution of a single stellar population of the outer stellar halo ($ R_{GC} > 15$ kpc) by an axisymmetric distribution following a single or a broken-double power law, depending of the complexity of the model. As we will soon see, a single power law is sufficient to provide an adequate description of the spatial distribution of our BHB sample. The generic form of this profile is given by:

\begin{equation}
\rho(m) = \rho_\odot \, (R_\odot/m)^{\gamma} \, ,
\end{equation}
where $\gamma$ is the slope of the power law and $\rho_\odot$ is the density of stars at the Solar radius ($R_\odot$). As we are interested only in the shape of the profile of the stellar halo traced by the BHB and not on the total number of BHB, $\rho_0$ is fixed to $1$ in our model.

Some recent studies favour a broken power law to model the smooth profile of the halo \citep[e.g. ][]{watkins_2009,deason_2014, xue_2015}, with a break radius around or below 20 kpc. To compare our result with these previous studies, we also implement a broken power law profile. The generic form is given by: 

\begin{equation}
\rho(m) = \rho_\odot \,  \left\lbrace
  \begin{array}{ l }
 (R_\odot/m)^{\gamma} \textrm{\ \ \ \ \ \ \ \ \ \ \ \ \ \ \  for $m \leq r_b$ } \\
  (R_\odot/r_b)^{\gamma-\beta} \, (R_\odot/m)^{\beta}\textrm{ for $m > r_b$}
\end{array}
\right. ,
\end{equation}
where $\gamma$ and $\beta$ are the inner and outer slope, respectively, and $r_b$ is the break radius (the radius where the change of the slope occurs).

In these models, we assume that the flattening is constant and independent of the Galactocentric radius. However, as pointed out by \citet{preston_1991} using both BHB and RR Lyrae, the flattening of the stellar halo may vary with distance (they find that the flattening decreases with Galactocentric radius). This result has been confirmed by \citet{carollo_2007} and \citet{schonrich_2011}, who identify two structural components to the stellar halo, the inner halo, that they argue is formed by \textit{in situ} stars and has an oblateness $q \sim 0.6$  and the outer halo, that they argue is formed via accreted  stars, that is more spherical with an oblateness of $q = 0.9 - 1.0$. Following \citet{hernitschek_2018}, we implemented a variation of the flattening of the halo as a function of Galactocentric distance ($R_{GC}$) for the single power law profile such that:

\begin{equation}
q(R_{GC}) = q_{\infty}-(q_{\infty}-q_0) \, \exp  \left(1 - \frac{\sqrt{R_{GC}^2+r_q^2}}{r_q} \right)  \, ,
\end{equation}

\noindent where $q_0$ is the the flattening at the center of the halo, and $q_{\infty}$ is the flattening at large galactocentric distance. $r_q$ is a characteristic radius marking a change between these values. 

We did not implement a triaxial model since, as illustrated in Figure.~\ref{plot_lb}, a great fraction of the northern Galactic hemisphere is not observed by CFIS at the present time.

\subsection{Construction of the selection function}

A good estimate of the selection function is mandatory to account for the observational biases that can affect our estimate of the real shape of the stellar halo, such as the completeness of the BHB sample or the spatial footprint of the survey. We separate our selection function in different categories to take into account these different effects. 

First, the CFIS footprint leads us to use in our analysis only the region covered by the survey, such that :

\begin{equation}
 \mathcal{S}_{area} (l,b) =
  \left\lbrace
  \begin{array}{ l }
 1 \textrm{ if $(l,b)$ in CFIS} \\
 0 \textrm{ otherwise}
\end{array}
\right .
\end{equation}

Our study is focused on the profile of the outer stellar halo ($> 15$ kpc), and so we only use stars that we estimate lie at a Galactocentric distance between 15 and 220 kpc (corresponding to the distance of the farthest BHB in our sample), as described by : 

\begin{equation}
 \mathcal{S}_{outer \, halo} (R_{GC}) =
  \left\lbrace
  \begin{array}{ l }
 1 \textrm{ if $15 < R_{GC} < 220$ kpc} \\
 0 \textrm{ otherwise}
\end{array}
\right .
\end{equation}

We notice that some point sources, identified as BHB stars by our algorithm, vicinity of the to cluster in the Andromeda (M 31) and Triangulum (M33) galaxies, and trace the shape of these galaxies (see Figure~\ref{M31_M33}). At the distance of M 31, $778\pm 19$ kpc \citep{conn_2011,conn_2012}, a typical BHB star should have an apparent magnitude of $z =24.95$, much fainter than the detection limit of the PS1 data. Thus, these point sources are probably young ($< 10$ Myr) main sequence stars or even star clusters, which have an absolute magnitude of $M_g \sim -5$ \citep{davidge_2012}. Two known galactic objects, Draco, NGC 2419 and NGC 5466, are also present between $15< R_{GC}<220$ kpc in the CFIS footprint, and their presence would impact the determination of the radial profile of the smooth halo. Thus we remove five regions around M 31, M 33, NGC 2419, NGC 5466 and Draco though the selection function so that:

\begin{equation}
 \mathcal{S}_{conta} (l,b) =
  \left\lbrace
  \begin{array}{ l }
 0 \textrm{ if \ \ \ \ \ \ $d_{M31} < 4.0$ deg } \\
 0 \textrm{ if \ \ \ \ \ \ $d_{M33} < 2.0$ deg} \\
  0 \textrm{ if $d_{NGC 2419} < 0.4$ deg} \\
 0 \textrm{ if $d_{NGC 5466} < 0.4$ deg} \\
 0 \textrm{ if \ \ \ \ $d_{Draco} < 0.5$ deg} \\
 1 \textrm{ otherwise}
\end{array}
\right .
\end{equation}
where $d_{M31}$, $d_{M33}$, $d_{NGC 2419}$, $d_{NGC 5466}$ and $d_{Draco}$  are the angular separation of stars relative to the centers of M 31, M 33, NGC 2419, NGC 5466 and Draco respectively. 

 \begin{figure}
\centering
  \includegraphics[angle=0,  clip, width=7.5cm]{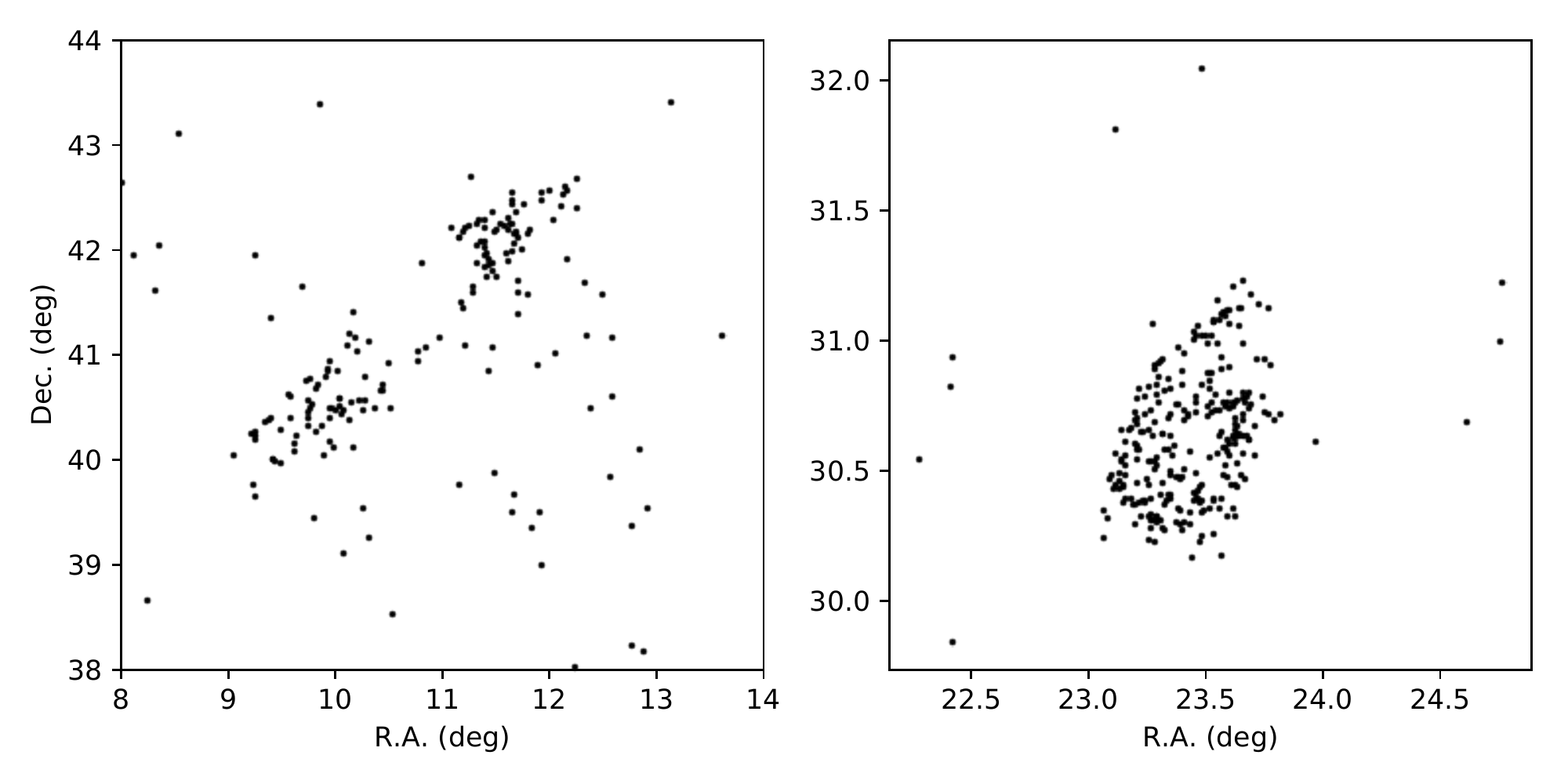}
   \caption{Point sources incorrectly identified as BHB stars around the Andromeda (left panel) and the Triangulum (right panel) galaxies.}
\label{M31_M33}
\end{figure}

As pointed out by \citet{deason_2011}, substructures -- and particularly the Sagittarius stream (Sgr stream) -- could affect the determination of the slope of the smooth halo profile. A significant portion of our survey contains the Sgr stream ($\sim 1/5$), and it is important to account for it. We prefer to remove all stars in the footprint that fall within 10 deg  \citep{majewski_2003} of the main Sgr stream orbit, rather than removing only the stars that have a good probability to be part of the stream using their distances. The latter method is dependent on a model of the variation of the distance to the stream, that is usually assumed to match the simulation of \citet{law_2010} (although this does not reproduce the distance of the farthest arm of the stream \citealt{belokurov_2014,thomas_2017}). The selection function we implement is thus given by :

\begin{equation}
 \mathcal{S}_{Sgr} (l,b) =
  \left\lbrace
  \begin{array}{ l }
 0 \textrm{ if $|\tilde{B}| < 10.0$ deg } \\
 1 \textrm{ otherwise}
\end{array}
\right .
\label{hide_sgr}
\end{equation}
\noindent where $\tilde{B}$ is the longitude of the Sgr stream coordinate system of \citet{belokurov_2014}. 

We also account for the completeness as a function of the magnitude. Here, extinction plays a role. We know the mean absolute magnitude in the $z$-band for our photometric BHB sample is $\langle M_{z,\textrm{BHB}}\rangle = 0.98$, and so it is possible to  calculate the mean apparent magnitude of a BHB at different distances and at different positions, so that:
\begin{equation}
  \left .
  \begin{array}{ l }
z_{\textrm{BHB}} (l,b,r_{helio}) = <M_{z,\textrm{BHB}}>  -\, 5 \\
+5\, \log( r_{helio} (1000.0/\textrm{kpc}))  +A_z \, ,
\end{array}
\right .
\label{z_bhb}
\end{equation}
\noindent where $A_z$ is the Galactic foreground extinction in the $z$-band and $z_{\textrm{BHB}}$ is the mean apparent magnitude for a BHB at a distance $ r_{helio}$.

Therefore, the selection function for the completeness of the BHB sample can be calculated from Equations~(\ref{comp_eq}), (\ref{lim_mageq}) and (\ref{z_bhb}) so that:
\begin{equation}
 \mathcal{S}_{comp} (l,b, r_{helio}) =  C_z \left( z_{\textrm{BHB}}- z_{lim}(l,b)+z_{lim,ref} \right)
\end{equation}

The overall selection function of our model accounting for the observations is given by:
\begin{equation}
\begin{split}
 \mathcal{S} (l,b,D) = &\mathcal{S}_{area} (l,b) \times  \mathcal{S}_{outer \, halo} (D)\\
 & \times  \mathcal{S}_{conta} (l,b) \times  \mathcal{S}_{Sgr} (l,b)\\
 & \times  \mathcal{S}_{comp} (l,b, D),
 \end{split}
\end{equation}
where $D$ is distance, either $r_{helio}$ or $R_{GC}$ depending on the term in the equation.

\subsection{Constraining the model}

With the selection function $ \mathcal{S}$, it is now possible to calculate the likelihood of the data given a set of parameters $\boldsymbol{\theta}$ for each of our three models of density profile $\rho_{\textrm{BHB}}(\mathcal{D} | \boldsymbol{\theta})$, defined in Section~\ref{profile_model}, in the same way as for \citet{hernitschek_2018}. The likelihood, $p_{\textrm{BHB}}(\mathcal{D}_i | \boldsymbol{\theta})$, of the $i$-th star, for a given profile of the BHB stars with the set of data $\mathcal{D}_i$, can be calculated as:

\begin{equation}
p_{\textrm{BHB}}(\mathcal{D}_i | \boldsymbol{\theta}) = \frac{\rho_{\textrm{BHB}}(\mathcal{D}_i | \boldsymbol{\theta})\, |\mathbf{J}|\, \mathcal{S}(l_i, b_i, D_i) }{\int \int \int \rho_{\textrm{BHB}}(l, b, D | \boldsymbol{\theta})\, |\mathbf{J}|\, \mathcal{S}(l, b, D)\, \textrm{d}l\, \textrm{d}b\, \textrm{d}D} \, .
\label{likelihood}
\end{equation}
The denominator of this equation is the normalization factor, where the integral is over the observed volume. As pointed out by \citet{hernitschek_2018}, the Jacobian term $|\mathbf{J}|= D^2 \cos b$ is required to transform from the Cartesian to Galactic coordinates.

As mentioned in Section \ref{BHBsel_par}, we estimate that up to 24\% of our photometric BHB sample may be contaminated from other A-types stars, mostly composed of BS. At a given distance, the BS population is less luminous than the BHB  \citep[$M_{g, BS} \simeq 2.5$ and $M_{g, BHB} \simeq 0.7$][]{deason_2011}. By misidentifying BS as BHB, we can potentially modify the derived profile, particularly at large radius since we will misidentify faint BS in the disk as bright BHB in the distant halo. To account for this contamination, we define the unmarginalized likelihood $p(\mathcal{D}_i | \boldsymbol{\theta})$ of the $i$-th star as: 

\begin{equation}
p(\mathcal{D}_i | \boldsymbol{\theta}, \boldsymbol{\theta}_{conta}) = (1- \alpha) \, p_{\textrm{BHB}}(\mathcal{D}_i | \boldsymbol{\theta} ) \, + \alpha \, p_{\textrm{conta}}(\mathcal{D}_i | \boldsymbol{\theta}_{conta})  \, ,
\label{likelihood_full}
\end{equation}

for a given BHB profile defined by the set of parameters $\boldsymbol{\theta}$ and a contamination profile defined by the parameter set $\boldsymbol{\theta}_{conta}$. $\alpha$ is the fraction of the sample due to contaminant stars, fixed at $\alpha = 0.24$. $p_{\textrm{conta}}(\mathcal{D}_i | \boldsymbol{\theta}_{conta})$ is the likelihood of the $i$-th star for a given contamination profile. This last term can be calculated with the same method used to calculate  $p_{\textrm{BHB}}(\mathcal{D}_i | \boldsymbol{\theta} )$ described by Eq. (\ref{likelihood}), replacing $\rho_{\textrm{BHB}}(l, b, D | \boldsymbol{\theta})$ by $\rho_{\textrm{conta}}(l, b, D | \boldsymbol{\theta}_{conta})$. The determination of the density distribution of the contaminant stars is detailed Section \ref{conta_sec}. 

The posterior probability of the set of parameters $\boldsymbol{\theta}$ is equal to $\ln p(\boldsymbol{\theta},\boldsymbol{\theta}_{conta}|\mathcal{D})  = \sum_{i}\ln p(\mathcal{D}_i | \boldsymbol{\theta},\boldsymbol{\theta}_{conta}) + p(\boldsymbol{\theta})$, where $p(\boldsymbol{\theta})$ is the uniform flat prior of the set of parameters. 

For the single power law profile with a constant oblateness, the parameters are defined over the following ranges:
\begin{equation}
 \left.
  \begin{array}{ l }
1.0 \leq \gamma \leq 6.0\\
0.1 \leq q \leq 2.0 \, ,
  \end{array}
  \right.
\end{equation}

For the single power law profile with $q(R_{GC})$, the parameters are defined over the following ranges: 
\begin{equation}
 \left.
  \begin{array}{ l }
  1.0 \leq \gamma \leq 6.0\\
15.0 \leq r_q (\textrm{kpc}) \leq 220.0\\
0.1 \leq q_0 \leq 2.0\\
0.1 \leq q_\infty \leq 2.0\, .
  \end{array}
  \right.
\end{equation}

Finally, for the broken power law profile, the parameters are defined over the following ranges: 
\begin{equation}
 \left.
  \begin{array}{ l }
  1.0 \leq \gamma \leq 6.0\\
   1.0 \leq \beta \leq 6.0\\
15.0 \leq r_q (\textrm{kpc}) \leq 220.0\\
0.1 \leq q \leq 2.0\, .
  \end{array}
  \right.
\end{equation}

To find the set of parameters that best match our data, we explore the parameter space with the Goodman \& Weare's Affine Invariant Markov Chain Monte Carlo \citep{goodman_2010} implemented by \citet{foreman-mackey_2013} in the Python module {\it emcee}. It is worth noting that from the initial $\simeq 10,200$ BHB stars in our sample, only $\simeq 5,900$ are in the outer stellar halo ($R_{GC} > 15$ kpc). Of these, $\simeq 1,100$ are in the Sgr regions. Thus, our study of the profile of the outer stellar halo is done using a sample of $\simeq 4,800$ BHB.

\section{Results \& Discussion} \label{results}

\subsection{The effect of the blue straggler contamination} \label{conta_sec}

To estimate the density distribution of contaminants, we assume that the normalized profile of the contamination is similar to the normalized profile of the stars that we identified as BS (Section \ref{BHBsel_par}), with distances derived under the assumption they are BHB. We refer to these stars as \textit{misidentified BS}). Figure \ref{profile_conta} shows that the profile of the BHB (assuming no contamination, black line) and the misidentified BS (gray line) have very different shapes. The number of BHB decreases rapidly after 100 kpc, due to the completeness of the sample. However, the number of BS decrease rapidly at a shorter distance, $\sim 70$ kpc. This is because most BS are located at much closer intrinsic distances (in the disk of the Galaxy), compared to the BHB that are mostly in the halo. 

We choose to model the density distribution of the misidentified BS by a broken power law.   We use the method described previously (with $\alpha=1$). The fitted profile is shown in Figure \ref{profile_conta} as a dashed red line (no selection effects) and as a solid red line (selection effects incorporated). As visible in Figure \ref{corner_conta}, the double broken power profile has an inner slope $\gamma= 2.95 \pm {0.03}$, an outer slope $\beta=4.03 \pm{0.06}$, a break radius $r_b = 73.7 ^{+ 2.9}_{-2.6}$ kpc and an flattening of $q = 0.58 \pm{0.01}$. We used this profile in Eq. (\ref{likelihood}) and  (\ref{likelihood_full}) to model the distribution of the contaminant stars present in our BHB sample, as described in the previous section.

 \begin{figure}
\centering
  \includegraphics[angle=0, viewport= 30 10 652 322,  clip, width=8cm]{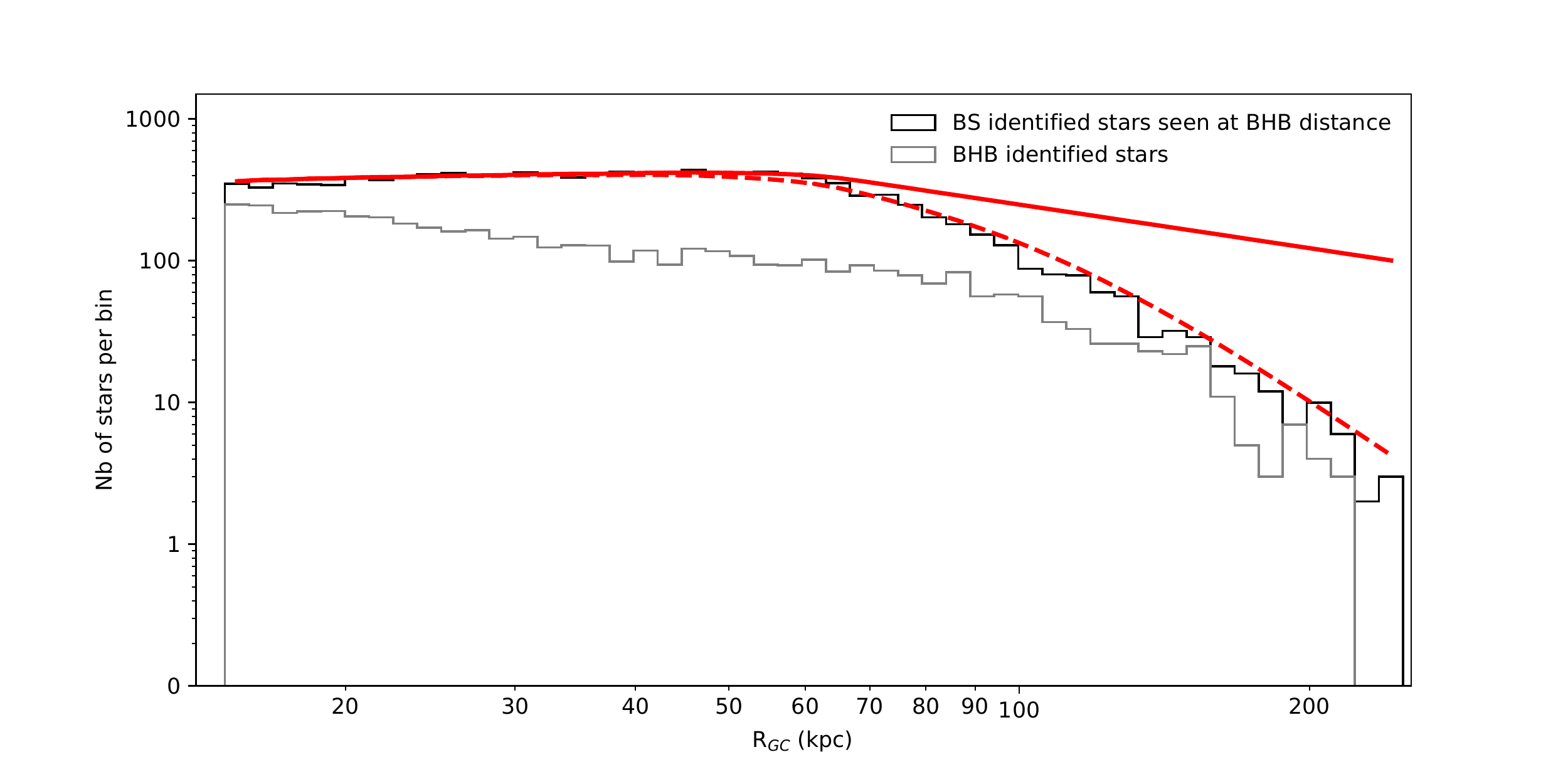}
   \caption{Comparison of the profile of the stars identified as BHB (in gray) with the profile of the stars identified as BS at the distance as if they were identified as BHB (in black). The solid red line and dashed red line shows the preferred broken power law fit to this profile as described in the text, with and without the selection effects, respectively.}
\label{profile_conta}
\end{figure}

\subsection{Results of the MCMC}

We apply the method described Section \ref{profile_model} on our BHB sample with the three density distributions mentioned previously. As illustrated by Figure~\ref{corner_PL}, in the case of the  single power law with a constant flattening, the distribution of BHB stars is best reproduced with a slope of $\gamma = 3.73^{+ 0.03}_{0.02}$ and a constant flattening of $q = 0.86 \pm 0.02$. The best-fit  parameters of the broken power law density profile has an inner slope $\gamma=4.24 \pm 0.08$, an outer slope $\beta=3.21\pm 0.07$, a break radius of $r_b=41.4^{+2.5}_{-2.4}$ kpc and a flattening $q=0.86 \pm 0.02$ (Figure~\ref{corner_BPL}). This is similar to the flattening found with the single power law. Finally, Figure \ref{corner_qvary} shows the best-fit parameters for the single power law model with a variation in the flatting as a function of radius. This model favors a stepper slope of $\gamma=3.89^{+0.06}_{-0.05}$ compare to model with a constant flattening. Furthermore, this model has an oblate central region with an inner flattening of $q_0 = 0.82\pm{0.02}$, a prolate shape in its outskirt with an outer flatting of $q_infty= 1.39^{+0.31}_{-0.19}$ and a transition radius of $r_q=119.9 ^{+48.0}_{-34.6}$ kpc. As shown by the inset panel in Figure \ref{corner_qvary}, this results in a halo that is oblate until $70$--$200$ kpc and prolate after this radius. The large uncertainties on the shape is mostly a consequence of the low precision with which we can measure the transition radius.

 \begin{figure}
\centering
  \includegraphics[angle=0, viewport= 0 0 723 728,  clip, width=8cm]{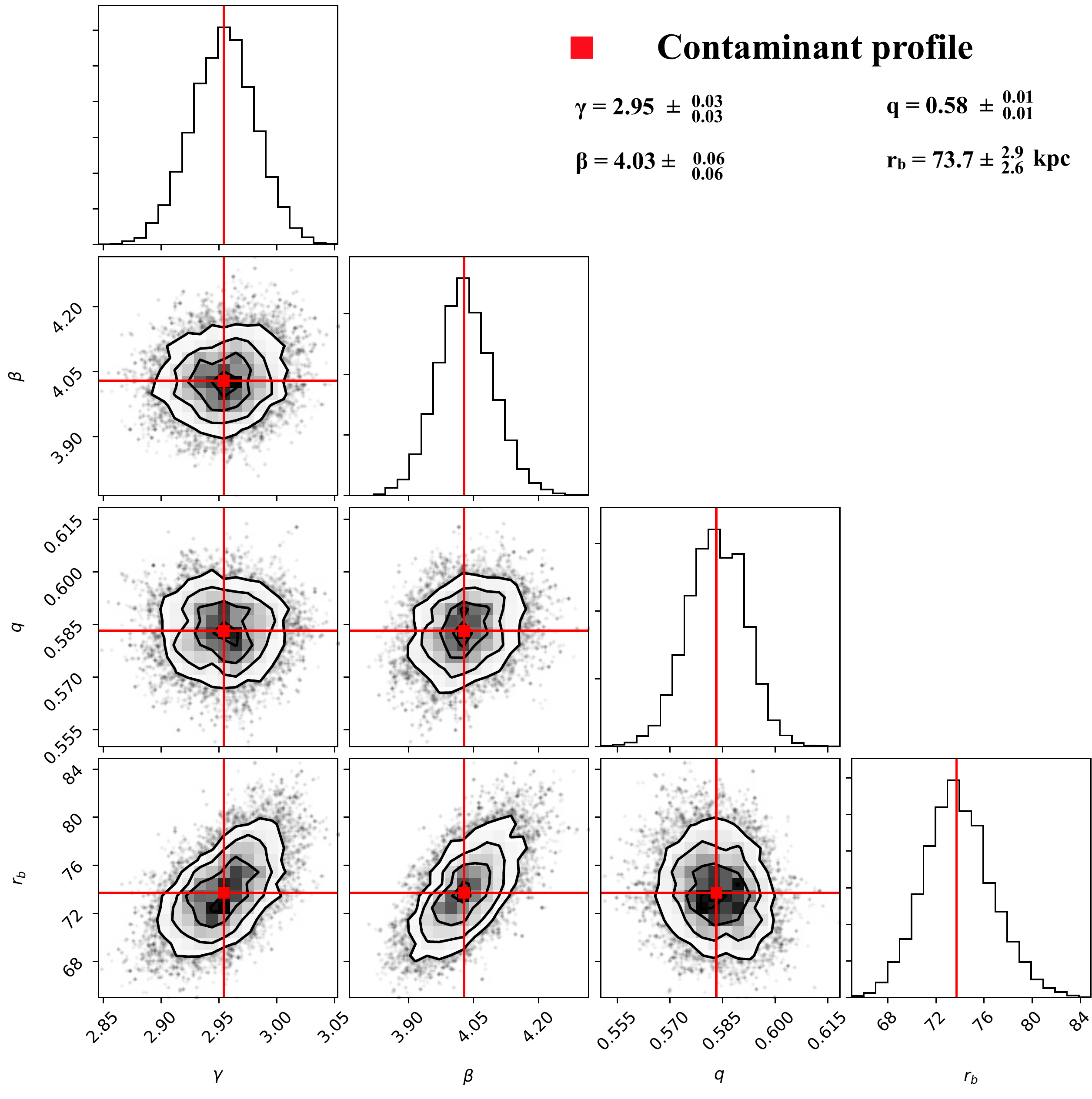}
   \caption{1-D and 2-D posterior distribution function of the parameters used to model the contamination distribution, assuming that the contamination follow a broken power law.}
\label{corner_conta}
\end{figure}

 \begin{figure}
\centering
  \includegraphics[angle=0, viewport= 0 10 621 623,  clip, width=8cm]{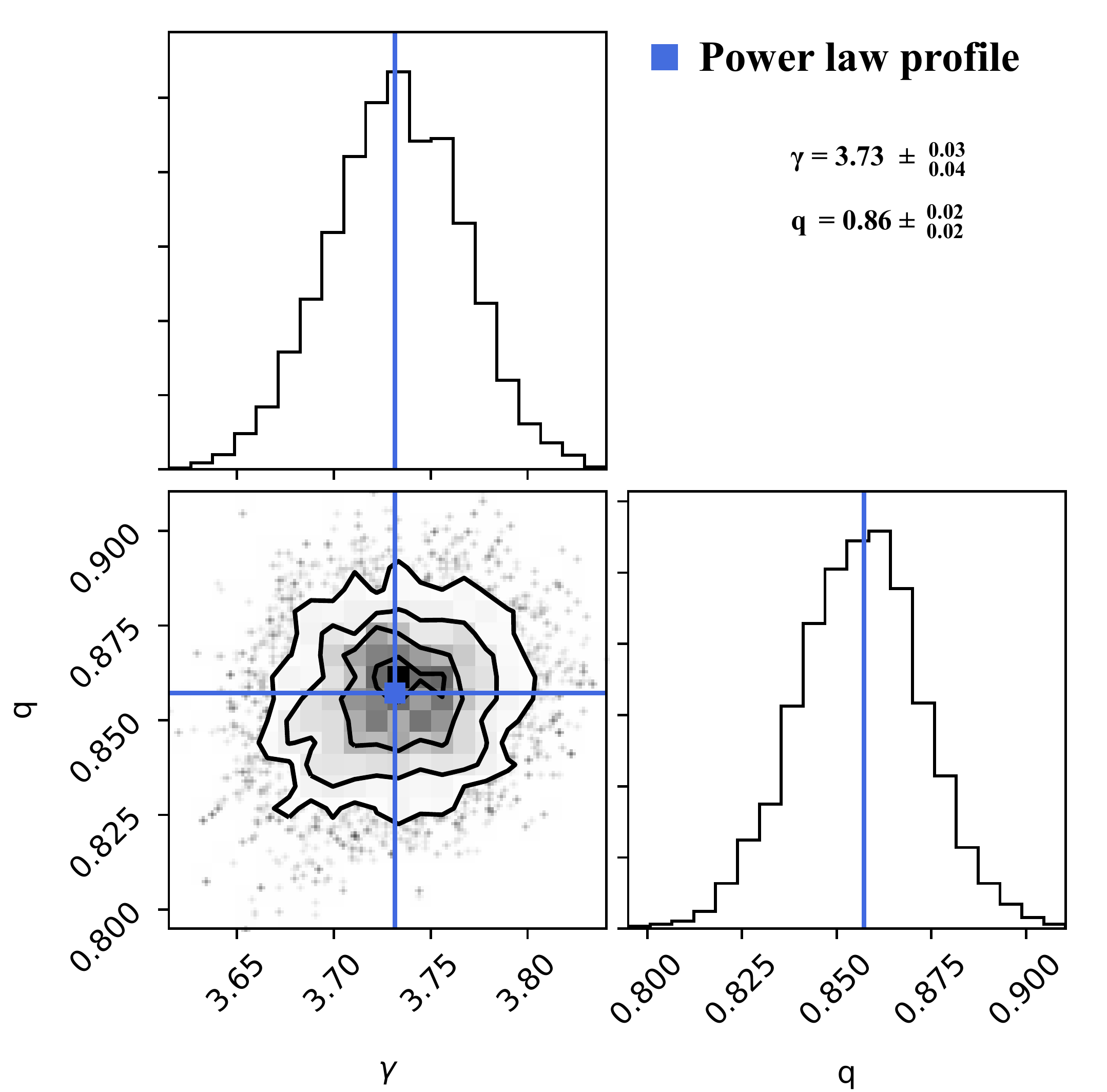}
   \caption{1-D and 2-D posterior distribution function of the parameters used in the single power law model with a constant flattening.}
\label{corner_PL}
\end{figure}

 \begin{figure}
\centering
  \includegraphics[angle=0, viewport= 0 0 777 723,  clip, width=8cm]{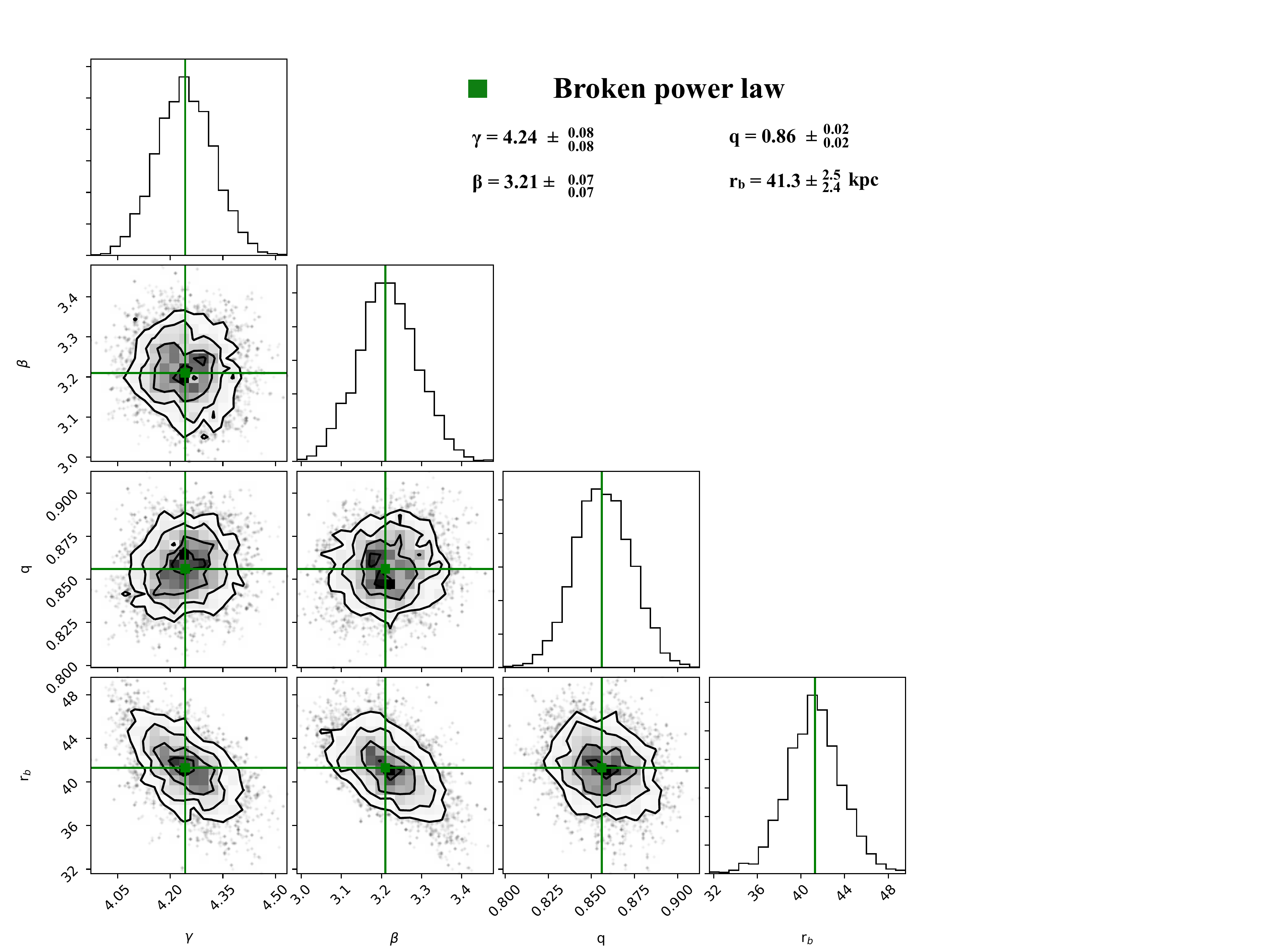}
   \caption{1-D and 2-D posterior distribution function of the parameters used in the broken power law model.}
\label{corner_BPL}
\end{figure}

 \begin{figure}
\centering
  \includegraphics[angle=0, viewport= 0 10 830 729,  clip, width=8.0cm]{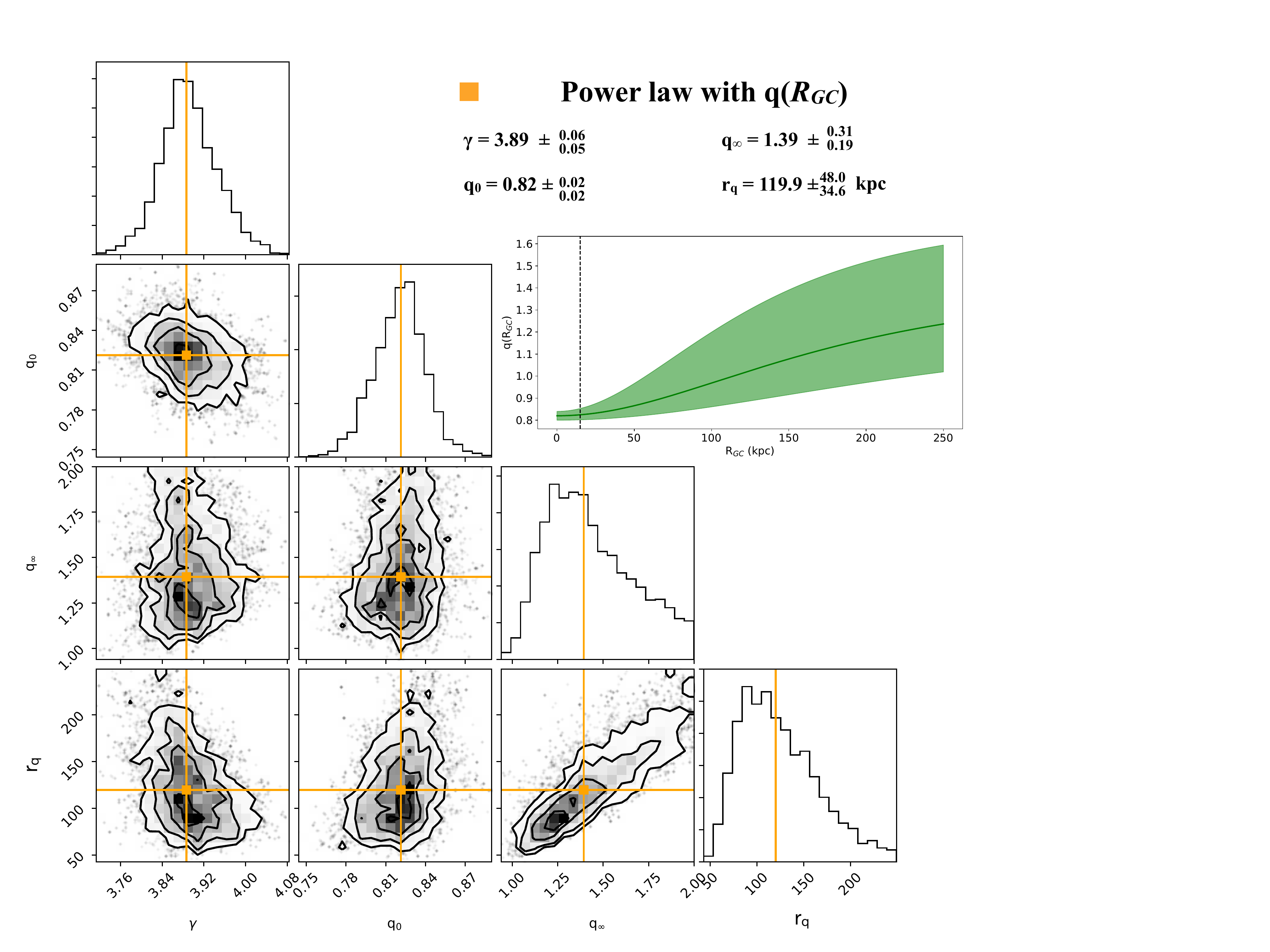}
   \caption{1-D and 2-D posterior distribution function of the parameters used in the single power-law model with a varying flattening.}
\label{corner_qvary}
\end{figure}

The best-fit versions of the distribution of BHB stars for each of these model are shown by the light blue, green and orange lines on Figure \ref{profile_plot} (single power law, broken power law, variable flatenning, respectively). The dashed red line shows the expected contamination from BS stars as discussed earlier. The observed radial profile of BHB stars is shown by the black histogram and the grey histogram show the total distribution of the BHB including the region covered by the Sgr stream, where a clear overdensity can be seen between $70 < R_{GC} < 90$ kpc, in agreement with the distance to the Sgr stream found by previous work in that region \citep{majewski_2003,koposov_2010,belokurov_2014,hernitschek_2017}. 

 \begin{figure*}
\centering
  \includegraphics[angle=0,  clip, width=18cm]{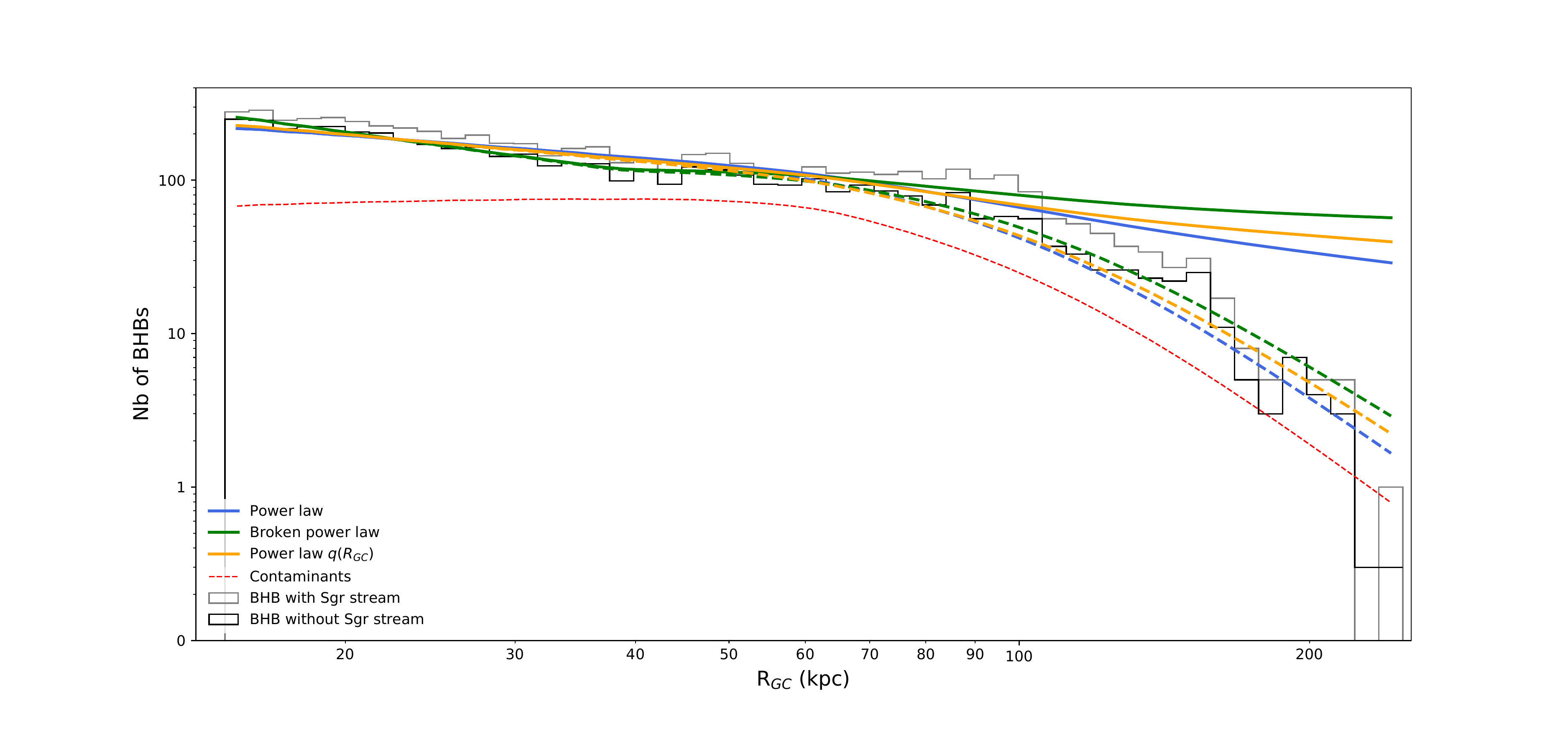}
   \caption{Number of BHB stars per distance interval. The black histogram shows our photometric BHB sample, excluding the Sgr region as described in the text. The gray histogram shows the same distribution, including the Sgr stream region. The light blue, green and orange curves show the predicted distribution of stars for our best-fit models using a single power law profile, the broken power law profile and the single power law with a varying flattening, respectively. The dashed lines show the same functions incorporating the observational biases encoded in the selection function. The expected contamination from BS is shown by the dashed red lines. }
\label{profile_plot}
\end{figure*}

\subsection{Preferred models}

To know which of our different models is statistically preferred, given the different number of parameters in each, we use the Bayesian information criterion. This is defined as:

\begin{equation}
\textrm{BIC}=\textrm{dim}(\theta) \, \ln(N) - 2 \ln{p(\mathcal{D} | \boldsymbol{\theta})_{max}} \, ,
\end{equation}
where $N$ is the number of BHB stars in our sample ($\sim 4,800$) and $\textrm{dim}(\theta)$ is the number of dimension of $\theta$, such that $\textrm{dim}(\theta)=2$ for the single power law profile with a constant flattening and $\textrm{dim}(\theta)=4$ for the other two models.

\begin{table}
 \centering
  \caption{Table of the BIC for each of our models. $\Delta$BIC refer to the difference between the BIC of a given model to the favorite model (with the lower BIC).}
  \label{BIC_table}
  \begin{tabular}{@{}lccc@{}}
  \hline
   Model & $\ln{p(\mathcal{D} | \boldsymbol{\theta})_{max}}$ & BIC & $\Delta$ BIC  \\
    \hline
  Power law & -23046 & 46109 & 58 \\
  Broken power law & -23008 & 46051 &  0\\
  Power law $q(R_{GC})$  & -23421 & 46877 & 826 \\
\hline
\end{tabular}
\end{table}

The different value of the BIC for the three models used in our study are given Table \ref{BIC_table}.  The broken power law is formally preferred, since a model with $\Delta$ BIC $>10$ indicates strong evidence against this model \citep{kass_1995}. Both models that adopt a constant flattening are strongly preferred over the model with variable flattening. 

 \begin{figure*}
\centering
  \includegraphics[angle=0,  clip, width=18cm]{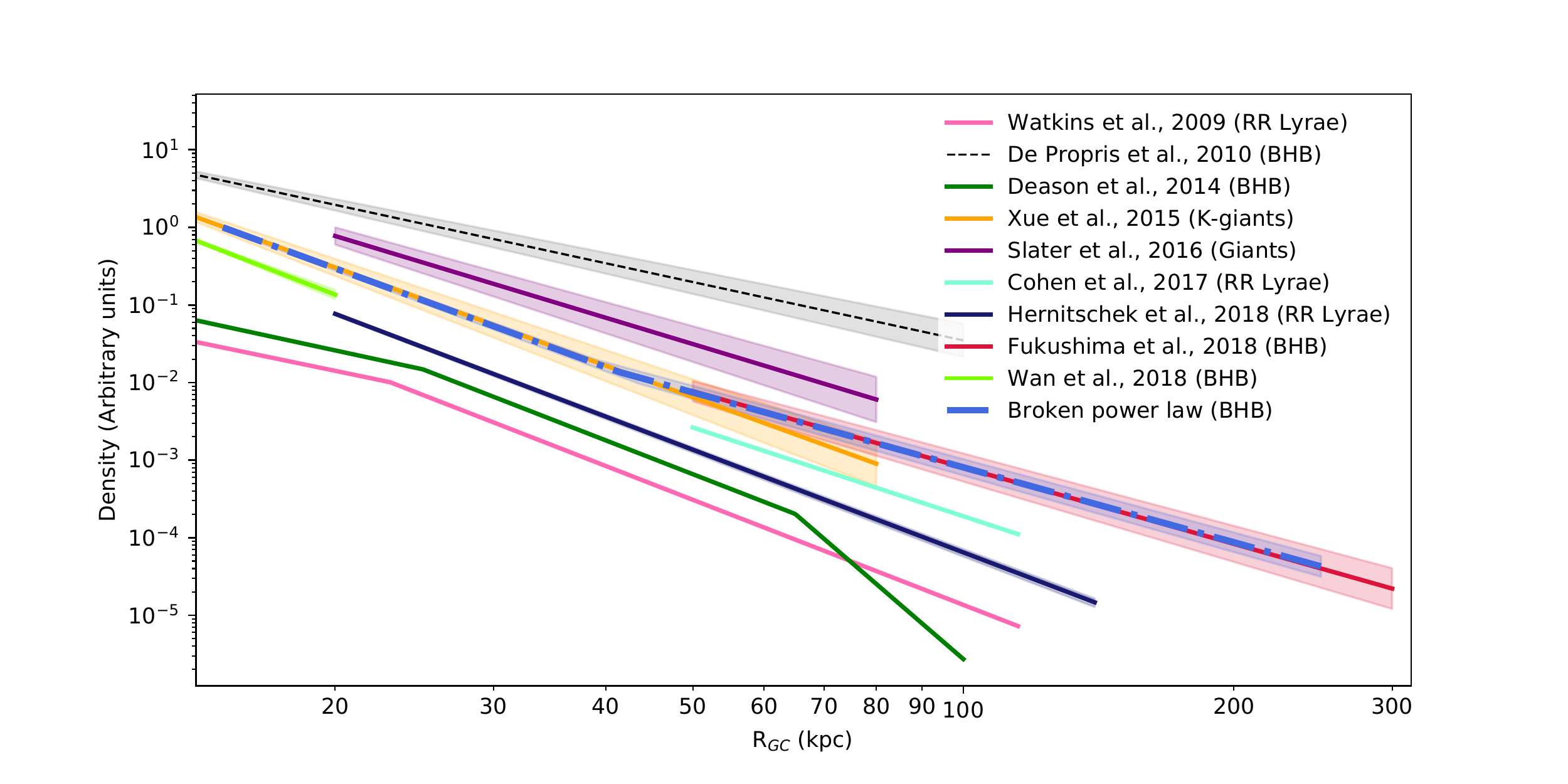}
   \caption{Comparison of our best-fit model of the power-law with a constant flattening (blue dot-dashed line) to best-fit models of other work. The slope of the stellar halo found by \citet{depropris_2010} using BHB stars is inconsistent with the slope found by other groups, including those studies that use the same tracer population.}
\label{profile_compare}
\end{figure*}

\subsection{Comparison to the literature}

Our analysis favors a broken power law profile, with an inner slope that is stepper than in the outskirt of the halo, with a transition around 40 kpc. The difference between the slope in the outer and inner region is $\sim -1$. This is similar to the difference found by \citet{hernitschek_2018} with the RR Lyrae from PS1, and these authors also  found a break radius around 40 kpc. However, their analysis favor a single power law profile and the absolute values of their inner and outer slope are much steeper that the slope found using the CFIS BHB stars. 

The inner slope ( $\gamma=4.24 \pm 0.08$) is close to the recent measure of $4.5 \pm 0.3$ between 11.8 and 20 kpc done by \citet{wan_2018} using BHB from SkyMapper. Our measurement is also similar than the slope of 4.5 found by \citet{watkins_2009} and than the single power law profile of \citet{hernitschek_2018} with a slope of $4.40^{+0.05}_{-0.04}$ with RR Lyrae. Moreover, our inner slope is in agreement with the slope of $\gamma=4.5$ found by \citet{deason_2014} using BHB stars in the SDSS between 25 and 65 kpc. However \citet{deason_2014} find a very steep slope of $6 - 10$ at large distances. We postulate that this very steep slope is possibly a consequence of an incorrect estimate of the  completeness of their BHB sample. For example, we can see in Figure~\ref{profile_plot} that the observed profile of the BHB sample is much steeper after $\sim 90$ kpc, but this change in slope is fully accounted for by the completeness correction. After the break radius located at $r_b=41.4^{+2.5}_{-2.4}$, the BHB profile is shallower and has an outer slope of $\beta=3.21\pm 0.07$. This is consistent with the slope of $3.2$ found \citet{fukushima_2018} between 50 and 210 kpc but is more shallower than the slope found by \citet{watkins_2009,cohen_2017,hernitschek_2018} after 50 kpc with the RR Lyrae. All these profiles are much steeper than the value of $\gamma=2.5$ found by \citet{depropris_2010} with the Two-Degree Field Quasar Redshift Survey; the slope of this profile is more than 30 $\sigma$ away from our measurement.

Other tracers have also been used to trace the profile of the outer stellar halo and these are summarized in Figure~\ref{profile_compare}. \citet{bell_2008} show that the profile of the stellar halo can be described by a power law slope of $2 - 4$ based on a sample of over 4 million main sequence turn-off stars out to 40 kpc. \citet{pila-diez_2015}, using F-stars and find a steep slope of $4.85$ out to 60 kpc. 
\citet{slater_2016}, using a sample of photometricly selected giants from DDO 51 and SDSS, find an index of $3.5$ up to $80$ kpc. \cite{xue_2015}, using K-giants from Segue, find a power law with an index of $4.2$ out to 80 kpc. We note that this last measurement is quite close to our estimate until 40 kpc. As discussed by \citet{hernitschek_2018}, it is difficult to determine if this difference between tracers are the consequence of intrinsic differences between the distribution of these different stellar populations, or if they are due to a difference in the methodology.

According to all these measurements, the outer stellar halo of the Milky Way is steeper than that for the Andromeda galaxy. \cite{ibata_2014b} find that the three dimensional density profile of M 31 is well reproduced by a spherical halo ($q=1.09$) with a single power-law index of $\gamma = 3.08$ for the old metal-poor red giant branch stars out to $\sim 300$ kpc. It is tempting to argue that this implies that the Milky Way is less massive, with a more quiet accretion history that has been contributed to by a lower number of large mergers than its neighbor M 31 \citep{bullock_2005,pillepich_2014,pillepich_2018}. However, it is risky to make broad statements on the history of accretion of these two galaxies with only consideration given to the slope of the outer halo. Indeed, the apparent inconsistency of the slope of the Milky Way stellar halo between different stellar populations necessitates a much more rigorous analysis. It will be interesting to explore these differences further, for example by using a code to generate synthetic stellar populations such as {\it Galaxia} \citep{sharma_2011} or the  Galactic Besan\c{c}on Model \citep{robin_2003}. It would also be interesting to compare these results with Milky-Way like galaxies in high resolution cosmological simulations such as {\it Auriga} \citep{grand_2017,grand_2018}.

\section{Summary} \label{conclusions}

In this paper, we use the new CFIS-$u$ survey in combination with PS1 to present a new photometric method to identify Blue Horizontal Branch stars (BHB). This new method reduces contamination from Blue Stragglers by a factor of 1.8 while having a completeness that is at least 1.2 times better than previous methods \citep{bell_2010,vickers_2012}. We study the completeness of our BHB sample as a function of magnitude and position, and show that our analysis is limited by the depth of the PS1 $z$-band data. 

We use the fact that BHB stars have a well constrained absolute magnitude \citep{deason_2011} to determine the profile of the outer smooth stellar halo up to a Galactocentric radius of $\sim 220$ kpc. We find that the outer stellar halo from 20 to 220 kpc is well reproduced by a broken power law with an inner slope of $\gamma=4.24 \pm 0.08$, an outer slope $\beta=3.21\pm 0.07$ after a radius of $r_b=41.4^{+2.5}_{-2.4}$ kpc, and a flattening $q=0.86 \pm 0.02$, close to spherical. This profile is in agreement with the recent measurement of \citet{fukushima_2018} who use BHB stars identified in the HSC-SSP, and with the study of \citet{wan_2018} who use BHB identify in SkyMapper. Although our inner slope is in agreement with the profiles traced by the RR Lyrae \citep{watkins_2009,cohen_2017,hernitschek_2018}, the profile of the stellar halo trace by the BHB beyond $\sim 40$ kpc is significantly shallower than determined with the RR Lyrae, that favor a steep single power law profile.

The variation of the halo profile as a function of stellar populations should be studied further in the future using synthetic stellar populations incorporated into cosmological simulations, to understand if this difference is a consequence of the method to select these different stellar populations, or if it is due to a physical effect. Moreover, in most simulations, the stellar halo has a steeper outer profile than is observed with the BHB \citep{bullock_2005,pillepich_2014,pillepich_2018,monachesi_2018}. 

We note that the shallower outer slope that we observe could be a consequence of a poor estimation of the contamination of our BHB sample by the BS. However, we consider this explanation  unlikely, since our outer slope is in agreement with the slope found by \citet{fukushima_2018} who use a different method to disentangle the BHB and the BS. Another explanation of a steeper inner slope than the outer slope is that a major merger have let more material in the inner region of the halo that in the outskirt and that the break detected in the BHB profile, that are old stellar population, are the imprint of an old major merger 8-11 Gyr ago as recently proposed by \citet{belokurov_2018a} with the Gaia data.

\section*{Acknowledgements}
This work is based on data obtained as part of the Canada-France Imaging Survey, a CFHT large program of the National Research Council of Canada and the French Centre National de la Recherche Scientifique. Based on observations obtained with MegaPrime/MegaCam, a joint project of CFHT and CEA Saclay, at the Canada-France-Hawaii Telescope (CFHT) which is operated by the National Research Council (NRC) of Canada, the Institut National des Science de l'Univers (INSU) of the Centre National de la Recherche Scientifique (CNRS) of France, and the University of Hawaii. This research used the facilities of the Canadian Astronomy Data Centre operated by the National Research Council of Canada with the support of the Canadian Space Agency.

We thank the anonymous referee for their careful reading and for their
helpful and constructive comments.

R. A. Ibata and N. F. Martin acknowledge support by the Programme National Cosmology et Galaxies (PNCG) of CNRS/INSU with INP and IN2P3, co-funded by CEA and CNES. This work has been published under the framework of the IdEx Unistra and benefits from a funding from the state managed by the French National Research Agency as part of the investments for the future program.

E. S. gratefully acknowledges funding by the Emmy Noether program from the
Deutsche Forschungsgemeinschaft (DFG).

\bibliography{./biblio}

\end{document}